\documentclass[11pt,twoside]{article}
\usepackage{epsf}
\newcommand{\etal}{{\em et al.{}}}
\newcommand{\be}{\begin{equation}}
\newcommand{\ee}{\end{equation}}
\newcommand{\bea}{\begin{eqnarray}}
\newcommand{\eea}{\end{eqnarray}}
\newcommand{\lapp}{\mbox{\raisebox{0.5ex}[-0.5ex][0mm]
                  {$\scriptstyle <$} \hspace{-1.3em}
    \raisebox{-2ex}[0mm][-2ex]{\Large \symbol{126}}}}
\newcommand{\gapp}{\mbox{\raisebox{0.5ex}[-0.5ex][0mm]
                  {$\scriptstyle >$} \hspace{-1.3em}
    \raisebox{-2ex}[0mm][-2ex]{\Large \symbol{126}}}}
\newcommand{\PostScript}[6]{
\begin{figure}[h]
\vspace{#2cm}
\begin{center}
\epsfysize=#1cm
\leavevmode
\epsfbox{#3}
\par
\end{center}
\mycaption{#4}{figure}{#5}{#6}
\end{figure}
}
\newcommand{\mycaption}[4]
              {\begin{center}
               \vspace{#1cm}
               \parbox{8.6cm}{
                \caption[#2]{\renewcommand{\baselinestretch}{1} \small \em #3}
                \label{#4}
               }
               \end{center}
              }

\begin{document}
%**************************************************************************
%                        Hier geht´s los
%**************************************************************************
%\onecolumn
%--------------------Titelseite-------------------------------
\rightline{\Large \bf TPR-96-20}
\vspace{0.5cm}
\rightline{Phys. Rev. A{\bf 56}, 182 (1997)}
\vspace{3cm}
\begin{center}
 {\Huge  Periodic orbit theory of a \\
           circular billiard \\
        in homogeneous magnetic fields \\}
\bigskip
\bigskip
\Large
     J. Blaschke and M. Brack \\
     Institut f\"ur Theoretische Physik \\
     Universit\"at Regensburg\\
     93040 Regensburg, Germany\\
\end{center}
\bigskip
\bigskip

\begin{center}
 \begin{minipage}{15cm}
  \small
  We present a semiclassical description of the level density of a
  two-dimensional circular quantum dot in a homogeneous magnetic
  field. We model the total potential (including electron-electron
  interaction) of the dot containing many electrons by a
  {\em circular billiard}, i.e., a hard-wall potential. Using the extended
  approach of the Gutzwiller theory developed by Creagh and
  Littlejohn, we derive an analytic semiclassical trace formula. For
  its numerical evaluation we use a generalization of the common
  Gaussian smoothing technique. In strong fields {\em orbit bifurcations},
  boundary effects ({\em grazing orbits}) and
  diffractive effects ({\em creeping orbits}) come into play, and the
  comparison with the exact quantum-mechanical result shows major
  deviations. We show that the dominant corrections stem from
  {\em grazing orbits}, the other effects being much less important. We
  implement the boundary effects, replacing the Maslov index by a
  quan\-tum-\-mechani\-cal {\em reflection phase}, and obtain a good
  agreement between the semiclassical and the quantum result for all
  field strengths. With this description, we are able to explain the
  main features of the gross-shell structure in terms of just one or
  two classical periodic orbits.
 \end{minipage} 
\end{center}
{\bf pacs} \hspace{4mm}
           03.65.Sq, % Semiclassical theories and applications
           73.20.Dx, % Electron states in low-dimensional structures
           73.23.Ps  % Mesoscopic systems: other electronic properties

\twocolumn
%-------------------Introduction---------------------
\section{Introduction}
 \label{intro}
 The two-dimensional free-electron gas (2DEG) that occurs at the interface
 of suitably designed semiconductor heterojunctions (see, e.g.~\cite{JJharris})
 has attracted a lot of interest in the last years. This is partly due to the
 two dimensionality, which gives rise to new physical effects 
 (such as the quantum
 Hall effect) and  partly to the extremely high mobility of the electrons,
 which comes from the absence of (scattering) donors or acceptors in the plane
 of the electron gas. The most attractive feature, however, is the great
 variability of these systems. With electron-beam lithography additional
 lateral constraints of the 2DEG  down to structures of some 10 $nm$ can be
 realized. This length is well below the typical phase coherence length and the
 electron mean free path (in GaAs  both of them can be of the order of some
 $\mu m$), and can even be comparable to the Fermi wavelength of the electrons
 (typically 40 $nm$ for GaAs), so that in such structures quantum confinement
 effects play an important role. These systems are therefore accessible on a
 quantum scale, opening up tremendous new possibilities in device design.

 Various approaches heave been used to model the 2DEG with and without
 additional lateral confinement. Quan\-tum-\-mechani\-cal calculations tend to be
 rather involved and even for the simplest systems numerically very demanding.
 Classical approaches have the severe drawback that they ignore quantum
 interference effects, and are therefore applicable only if the system
 dimensions are long compared to the mean free path and the phase coherence
 length. In the resulting gap of the theoretical description, semiclassical
 approaches appear very promising. They approximate quantum mechanics in such a
 way that the quantities involved can be interpreted {\em classically}, often
 in terms of the classical orbits in the system. They  combine the advantages
 of the classical description, especially its limited numerical demands, with
 the ability to reproduce quan\-tum-\-mechani\-cal interference effects. This makes
 semiclassical methods a very attractive tool for {\em mesoscopic} systems,
 i.e., systems with dimensions comparable to the phase coherence length and the
 mean free path. One of the most striking successes in recent years has been
 the explanation of conductance oscillations in superlattices, the Weiss
 oscillations~\cite{weiss:oszi}.

 In this paper, we consider the level density of a 2DEG confined by external
 electric fields to a circular domain, with an additional homogeneous magnetic
 field perpendicular to the plane of the 2DEG. When this quantum dot contains
 many electrons, the effective single-particle potential (i.e., the Kohn-Sham
 potential in the language of density functional theory, which contains the
 elec\-tron-\-elec\-tron interaction in the local density
 approximation) is Wood-\-Saxon-like, 
 with a flat region in the interior and a rather steep
 surface.\footnote{This has been shown in self-\-consis\-tent calculations for
         quantum dots~\cite{darnhofer} and is analogous to the situation in
         three-dimen\-sional metal clusters~\cite{mathias:cluster}.}
 The level density is not too sensitive to details of the potential edge, so
 that a circular disk with infinite reflecting walls, i.e., a {\em circular
 billiard}, is a realistic model.

 The level density itself is hard to access experimentally, but it enters in
 many observable quantities. Persson {\em et al.{}}~\cite{persson}, for example,
 consider a quantum dot that is connected by two point contacts to the
 surrounding 2DEG. They propose an approximation in which the conductivity of
 this system in weak external magnetic fields is proportional to the level
 density of the dot at the Fermi energy.\footnote{Recent exact
         quan\-tum-\-mechani\-cal calculations of the transport properties, however,
         show that this approximation is only valid for contacts at the
         opposite sides of the dot~\cite{berggren}.}
 Their measurements on a circular dot with about 1000 -- 1500 electrons in a
 homogeneous magnetic field show characteristic conductance oscillations that
 could be well explained qualitatively in a perturbative approach by Reimann
 {\em et al.{}}~\cite{Steffi:B}. They reproduce the oscillations in a simple and
 intuitive way by a few classical periodic orbits of the system and the flux
 enclosed by them. Because of its perturbative nature, this description only
 holds in weak fields. Another example where the level density enters
 observable quantities is the magnetization~\cite{ullmo,kaori}.

 For the interpretation of experiments on circular quantum dots, a
 semiclassical approximation of the level density with arbitrary field strength
 is desirable. Such a description in terms of classical orbits is also of
 theoretical interest. In the absence of a magnetic field the classical orbits
 consist of straight paths bouncing at the boundary (see Fig.~\ref{OrbitsB0}).
 They have a one-\-dimen\-sional degeneracy corresponding to the rotational
 symmetry of the system. In very strong fields, the confinement is negligible
 and the level density is dominated by the quantization of a free-electron gas,
 leading quan\-tum mechani\-cally to the Landau levels and described
 semiclassically by closed cyclotron orbits. These orbits do not touch the
 boundary and have a two-\-dimensional translational degeneracy. A unified
 semiclassical description thus has to include the transition between these two
 limiting cases, which includes changes of the topology and the degeneracy of
 the classical orbits. Treating these is of conceptual interest since both
 effects are well known to lead to divergences in semiclassical theories.

 We conclude this Introduction with a short outline of the paper. As a
 reference, we first present in Sec.~\ref{QMdisk} the quantum-mechanical
 solution for the circular billiard in homogeneous magnetic fields.
 Section~\ref{SCtools} gives a short introduction to semiclassical methods, and
 in Sec.~\ref{POdisk} we derive a semiclassical trace formula
 of the disk. We then compare its results to the quan\-tum-\-mechani\-cal ones for
 various field strengths. The agreement is good for very weak and for very
 strong fields, but the semiclassical approach rather appears to fail in the
 intermediate regime. The deviations are due to bifurcations of classical
 orbits, to diffraction effects, and to boundary effects. The latter give the
 largest contributions, and in Sec.~\ref{Boundary} we develop a simple
 approximation to include these effects in the trace formula. This corrected
 trace formula gives satisfactory results for all magnetic field strengths, and
 we give an intuitive interpretation of the  the level density in the various
 $B$-field regimes. The paper closes with a summary of the results and an
 outlook to further investigations.
%------------------- The quantum-mechanical solution----------------------
\section{The quantum-mechanical solution}
 \label{QMdisk}
 In the following, we will use normalized energies $\widetilde E$ in units of
 $\hbar^2 / 2 m R^2$ and normalized magnetic fields $\widetilde B$ in units of
 $\hbar / e R^2 $, where $R$ is the disk radius. 
 In these units, we have $\sqrt{\widetilde E}=kR$ and with
 the classical cyclotron radius $R_c=\hbar k / eB$, we get $R_c / R =kR /
 \widetilde B $.

 The exact quan\-tum-\-mechani\-cal solution for the circular billiard in homogeneous
 magnetic fields was presented by Geerinckx~\cite{Gee} and, using a different
 approach, by Klama and R\"o\ss ler~\cite{klama}. 
 The eigenenergies are given by the
 zeros of the confluent hypergeometric function $ _1 F_1$ as
 \be
  \widetilde E_{nl}=2 \widetilde B \cdot
                   \left( \alpha_{nl} + \frac{1+|l|}{2}+\frac{1}{2}
                   \right) \quad ,
 \ee
 where
 \be
  _1 F_1 \left(-\alpha_{nl};1+|l|;
                \frac{\widetilde{B}}{2}
         \right) =0
  \quad .
 \ee
 Here $n>0$ denotes the radial and $l$ the angular-momentum quantum number. The
 zeros of $ _1 F_1$ were determined numerically --- which is conceptually easy
 but requires a lot of numerical work. Figure~\ref{Spectrum} shows the
 well-known dependence of the eigenvalues $\widetilde E_{nl}$ on
 $\widetilde{B}$. One clearly sees how with increasing magnetic field the
 different states condense into the Landau levels (dashed lines).
%                  ------------------------
\PostScript{5.8}{0}{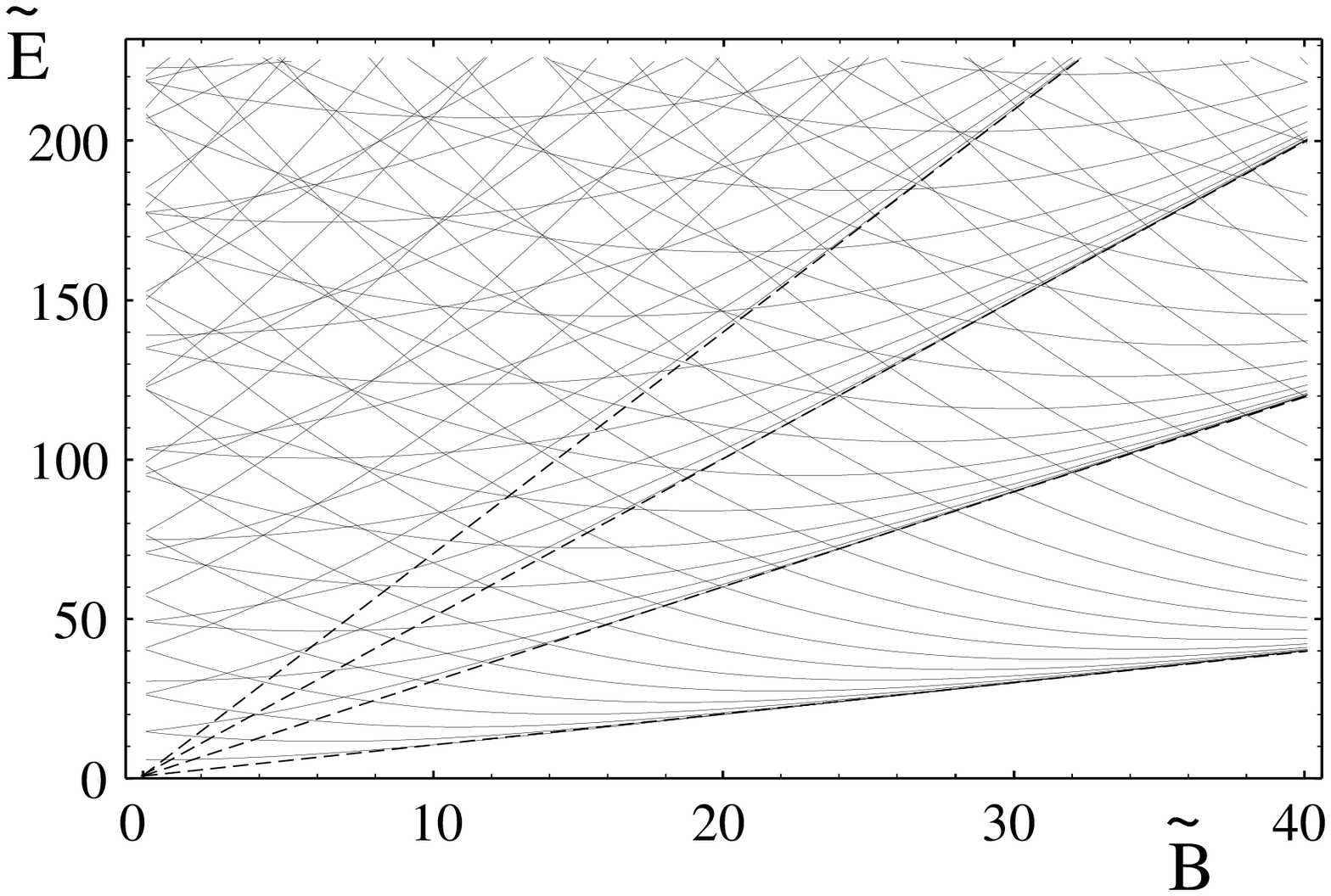}{-0.8}
           {The quan\-tum-\-mechani\-cal eigenenergies of the circular
           billiard in dependence of the magnetic field.
           The dashed lines correspond to the four lowest Landau levels.}
           {Spectrum}
%                  ------------------------

%-------------------Semiclassical methods---------------------
\section{Semiclassical methods}
 \label{SCtools}
 For semiclassical approximations, a broad variety of methods is at hand. Some
 of them approximate directly the quan\-tum-\-mechani\-cal eigenvalues
 \cite{WKB,EBK}, whereas others describe the level density $g(E)$. For this
 purpose, $g(E)$ is usually split up into a {\em smooth part} $\tilde{g}(E)$,
 the (extended) Thomas-Fermi level density, and an oscillating part $\delta g$:
 \begin{equation}
  g(E)=\tilde{g}(E) + \delta g(E) \quad.
 \end{equation}
 The latter can be expressed in terms of the periodic orbits of the
 corresponding classical system, which is therefore called 
 {\em periodic orbit theory} (POT). Such relations have been established by
 various approaches \cite{BerryTabor,BB2,gutz,sascha,steven:cont}, usually
 resulting in so-called {\em trace formulae} of the form
 \begin{equation}
  \label{trace_formula}
  \delta g(E) =  \frac{1}{\pi\hbar} \sum_{\Gamma} A_{\Gamma}
                 \sin\left(\frac{S_{\Gamma}(E)}{\hbar}
                           -\sigma_{\Gamma}\frac{\pi}{2}
                           +\frac{\pi}{4}
                     \right)
                 \quad .
 \end{equation}
 Here $\Gamma$ labels all classical periodic orbits of the system. Each orbit
 contributes to the level density via an oscillating term that depends on the
 classical action $S_\Gamma$ along the orbit and on the {\em Maslov index}
 $\sigma_\Gamma$ depending on the orbit's topology. The amplitude $A_\Gamma$ is a
 slowly varying function of energy, determined by classical properties of the
 orbit such as its degeneracy and its
 stability.\footnote{Especially, $A$ depends on $\hbar$ only by a
              factor $\hbar^{-k/2}$, where $k$ is the degree of
              degeneracy of the orbit family.}
 All semiclassical approaches have their individual
 merits and drawbacks~\cite{matthias:buch}, 
 and it is interesting to note that for the simple
 integrable case of the circular billiard without magnetic field, all
 applicable methods result in the same trace
 formula~\cite{tatievski,boga,mypaper}, which furthermore reproduces
 exactly~\cite{mypaper} the 
 EBK\footnote{EBK stands for the semiclassical approximation developed
              by Einstein, Brillouin and Keller~\cite{EBK}}
 spectrum. Applying a homogeneous
 magnetic field, the system remains integrable, with the energy and the $z$
 component of the conjugate angular momentum as constants of the motion. In
 weak fields, this system was treated using a perturbative approach by
 Bogachek and Gogadze~\cite{boga},
 Ullmo {\em et al.{}}~\cite{ullmo},
 and by Reimann {\em et al.{}}~\cite{Steffi:B}. We have
 chosen the trace formula of Creagh and Littlejohn~\cite{steven:cont} as a
 starting point for the description in arbitrarily strong fields.

%--------------------Trace formula for the disk billiard-------------------------
\section{Trace formula for the circular billiard}
 \label{POdisk}
 The trace formula of Creagh and Littlejohn~\cite{steven:cont} is well suited
 for the semiclassical description of the circular billiard, as it can deal
 with continuous symmetries. The main idea of their approach is the
 separation into a symmetry-free system treated by usual semiclassical
 techniques and the symmetry, which is used to integrate over the
 orbit families.
 The structure of the trace formula (\ref{trace_formula}) remains
 essentially unchanged by this procedure, but the definition of $A$ is
 different, reflecting the different classical structure of the
 dynamics. For the details we refer to the
 original publication~\cite{steven:cont}. In order to calculate the level
 density with this trace formula, we have to classify the periodic orbits and
 calculate their actions, amplitudes, and Maslov indices. These steps are
 presented in the following subsections.

%--------------------Classification of the periodic orbits----------------
 \subsection{Classification of the periodic orbits}
 \label{classification}
  The complete classification of the periodic orbits of our system is
  straightforward. Let us first consider the case without magnetic field. In a
  circular billiard, the periodic orbits (PO) are identical to those in a
  three-dimensional spherical cavity, 
  whose complete classification has been given
  by Balian and Bloch~\cite{BB2}. All orbits have a one-dimensional degeneracy
  corresponding to the rotational symmetry of the system. Each family of
  degenerate orbits with a given action (or length) can be represented by a
  regular polygon. The first few polygons are shown in Fig.~\ref{OrbitsB0}.
%                  ------------------------
\PostScript{4}{0.5}{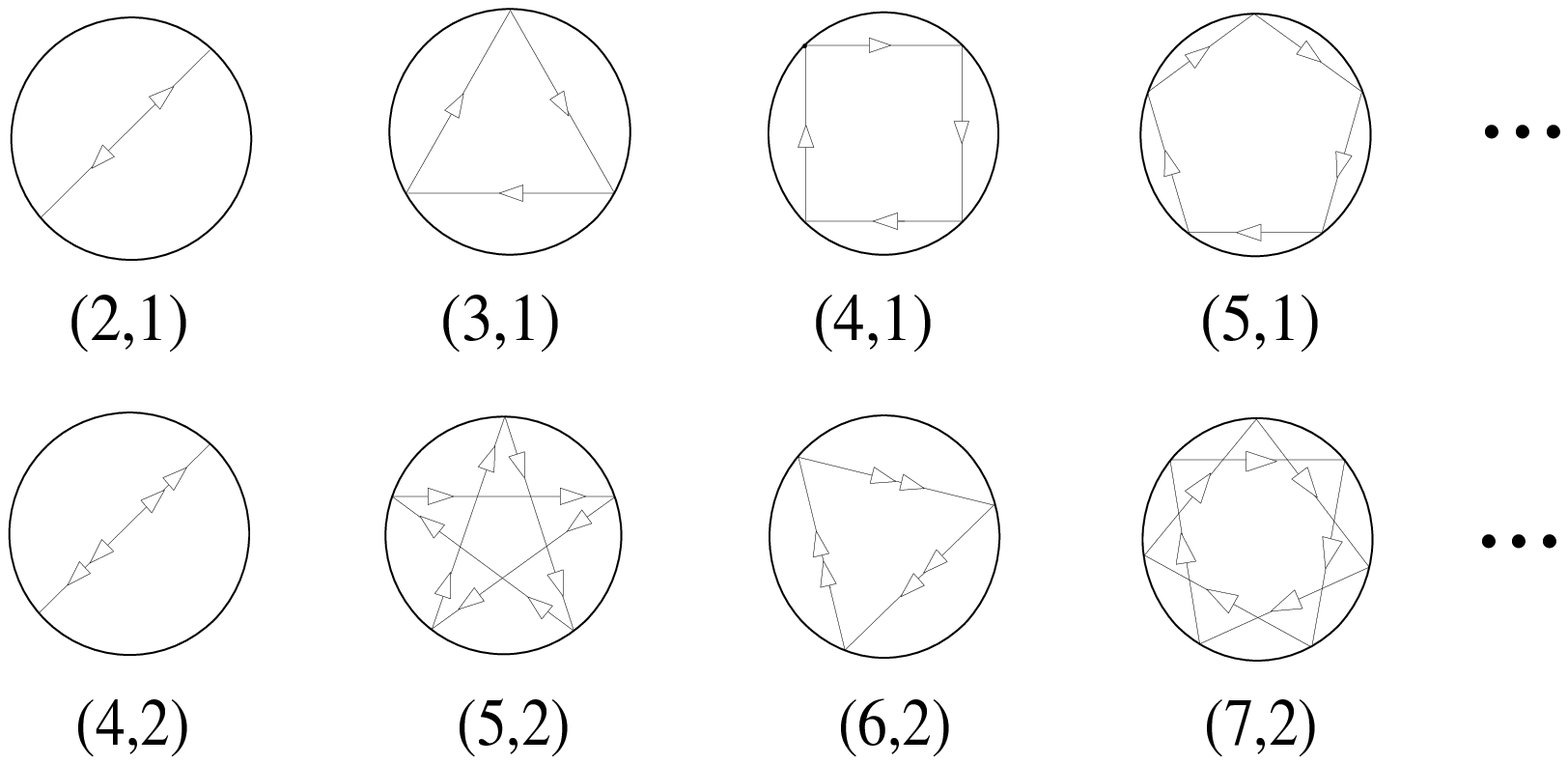}{-0.7}
           {The classical periodical orbits of the circular billiard
            in the absence of a  magnetic field are the regular
            polygons. They can
            be classified with $(v,w)$, where $v$ is the number of corners
            and $w$ indicates how often the trajectory winds round the
            center of the disk.}
           {OrbitsB0}
%                  ------------------------
  These orbit families are classified~\cite{BB2} by $\beta=(v,w)$, where $v$
  denotes the number of corners (vertices), and $w$ is the winding number,
  i.e., it counts how often an orbit winds around the center. 
  With $v \ge 2w > 2$, $\beta = (v,w)$ uniquely describes all families 
  of POs of the system in the absence of a magnetic field.
  Because of the
  time-reversal symmetry, all orbits except the diameter ($v=2w)$ have an
  additional discrete twofold degeneracy, which has to be accounted for in
  the trace formula.

  Switching on the magnetic field causes the classical trajectories to bend,
  with the direction of the curvature depending on the direction of motion with
  respect to the magnetic field. This entails a breaking of time-reversal
  symmetry. For weak fields, the orbits can still be classified by $\beta$ if
  an additional index ($\pm$) is introduced. This situation
  is shown in the upper row of diagrams in Fig.~\ref{OrbitsB} for the orbit
  $\beta=(4,1)$. Up to a field strength where the classical cyclotron radius
  $R_c$ equals the disk radius $R$, henceforth referred to as the 
  {\em weak-field regime}, 
  the orbits do not change their topology and the classification
  $\beta^\pm$ holds. For the {\em strong-field regime} with $\widetilde{B}>kR$,
  the structure of the POs is different. This situation is shown in the second
  row of diagrams in Fig.~\ref{OrbitsB}. The $\beta^-$ orbits change their
  shape continuously over the point $R_c=R$, but the $\beta^+$ orbits change
  their topology abruptly. However, since there is a one-to-one correspondence
  between orbits for $R_c \gapp R$ and for $R_c \lapp R$, $\beta^\pm$ still
  gives a complete classification of all {\em bouncing orbits}, i.e., of orbits
  that are reflected at the boundary. For $R_c<R$, there are additional {\em
  cyclotron orbits} that do not touch the boundary at all. They have to be
  included separately in the sum over all orbits in the trace formula.
%                  ------------------------
\PostScript{6}{0.3}{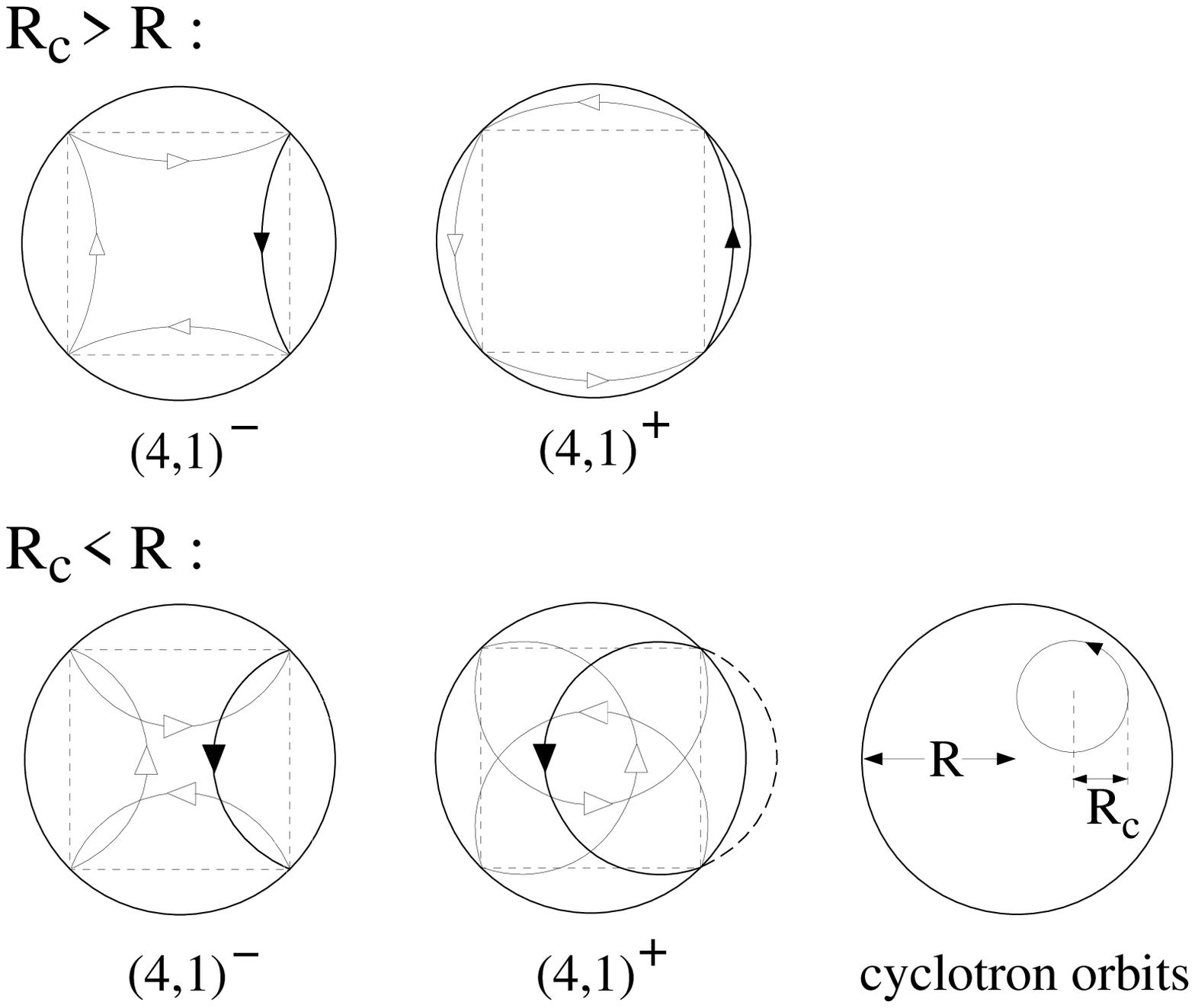}{-0.8}
           {A magnetic field breaks the time-reversal symmetry, so that the
           orbits are no longer independent of the direction of motion.
           Introducing an additional index $\pm$, the orbits can be classified
           by $(v,w)^\pm$, both in weak ($R_c>R$) and in strong ($R_c<R$)
           fields. For strong fields, there occurs an additional family of
           orbits that do not touch the boundary, the cyclotron orbits.}
           {OrbitsB}
%                  ------------------------
  At field strengths where $R_c \le R \sin (\pi w/v )$,
  the $(v,w)^\pm$ orbits no longer exist (see Fig.~\ref{bifurkation}). They
  vanish pairwise, which is the simplest case of an {\em orbit bifurcation}.
  This imposes an additional restriction on the sum over $(v,w)$. Including
  this, we now have a complete classification of all periodic orbits in the
  circular billiard at arbitrary field strengths.
%                  ------------------------
\PostScript{2.8}{0}{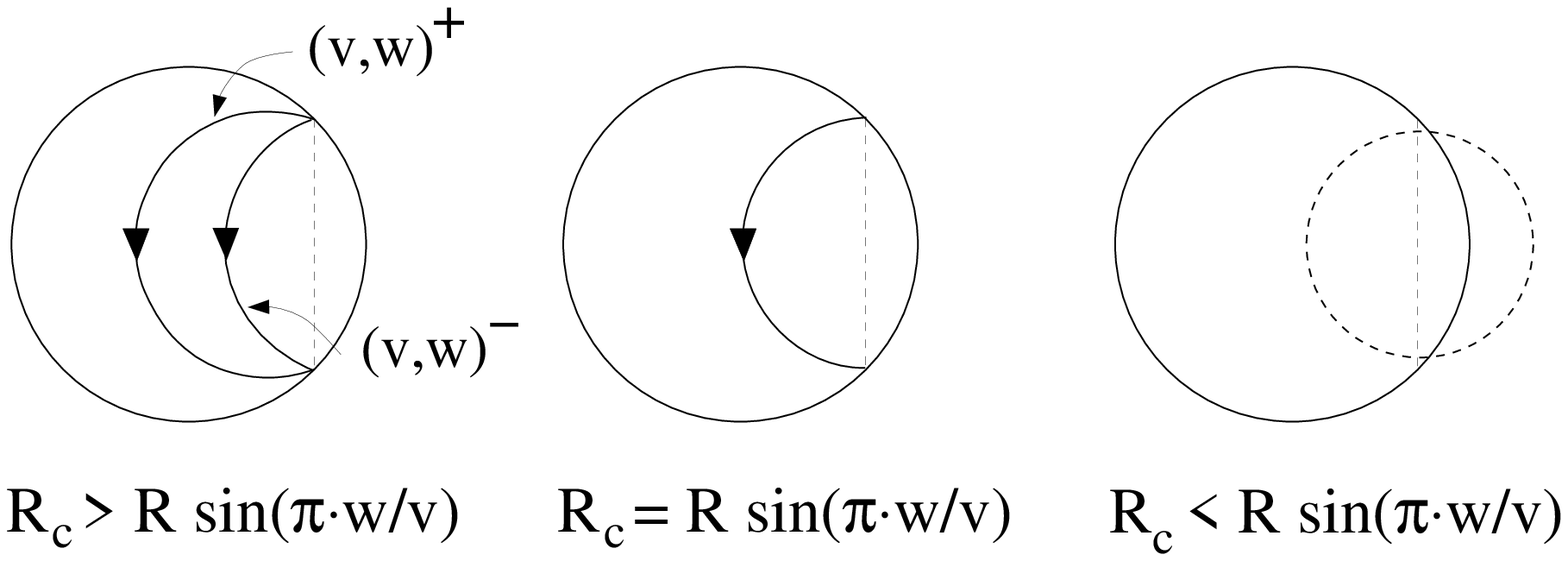}{-0.8}
           {At a field strength where $R_c=R \sin (\pi \; w/v )$, the orbits
           $(v,w)^\pm$ vanish pairwise.}
           {bifurkation}
%                  ------------------------
%----------------The bouncing orbits-----------------------------
 \subsection{The bouncing orbits}
  The action of a closed orbit in a magnetic field can be written as the sum of
  the kinetic part and the magnetic flux enclosed by the orbit
  \begin{equation}
    S_\beta= \int p \, \mbox{d} q = \hbar k L_\beta - e B F_\beta \qquad .
  \end{equation}
  The geometrical lengths $L_\beta$ and the enclosed areas $F_\beta$ of the
  periodic orbits discussed above (correctly counting those areas that are
  enclosed several times, cf.~Fig.~\ref{flux}) can be calculated by elementary
  geometry. In terms of the geometrical quantities $R_c,R,\gamma$ and $\Theta$,
  explained in Fig.~\ref{geoGR}, we obtain
%                  ------------------------
\PostScript{2.5}{0}{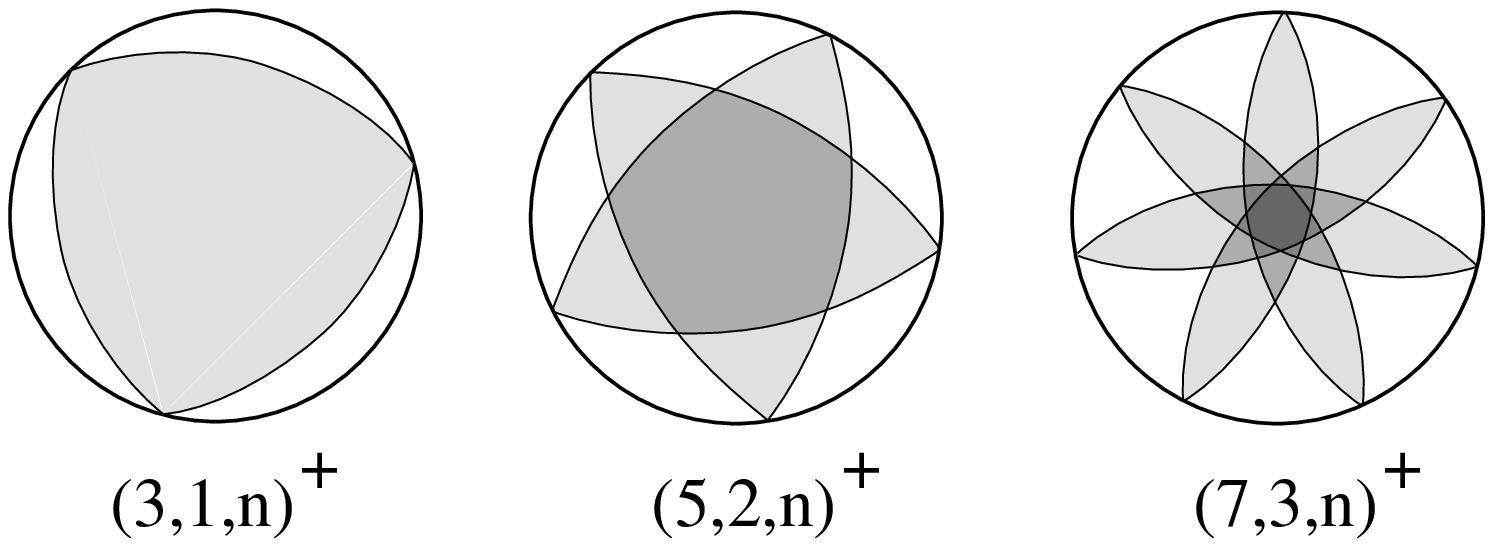}{-0.8}
           {Calculating the magnetic flux enclosed by an orbit, the
           multiply enclosed areas (darker gray) have to be correctly
           accounted for.}
           {flux}
%                  ------------------------
  \begin{eqnarray}
   \label{actions}
   S_\beta(E)&=&v \hbar kR_c\eta \quad ,   \\
                  \eta&=& \left\{
                          {\setlength{\arraycolsep}{0.1em}
%                           \begin{array}{rcccc@{\hspace{10mm}}l}
                           \begin{array}{rccccp{5mm}}
                             \gamma & + & \frac{\sin 2\gamma}{2} & - &
                               \left(\frac{R}{R_c}\right)^2
                                    \frac{\sin 2\Theta}{2} &
                               \\ \multicolumn{6}{r}
                                  {\rule{0mm}{2.5ex}
                                   \mbox{for } (\beta^+,R_c \ge R)}  \\
                             \rule{0mm}{4.5ex}
                             \pi-\gamma & - & \frac{\sin 2\gamma}{2} & + &
                               \left(\frac{R}{R_c}\right)^2
                                    \frac{\sin 2\Theta}{2}&
                               \\ \multicolumn{6}{r}
                                  {\rule{0mm}{2.5ex}
                                   \mbox{for } (\beta^+,R_c \le R)}  \\
                             \rule{0mm}{4.5ex}
                             \gamma & + & \frac{\sin 2\gamma}{2} & + &
                               \left(\frac{R}{R_c}\right)^2
                                  \frac{\sin 2\Theta}{2}&
                               \\ \multicolumn{6}{r}
                                  {\rule{0mm}{2.5ex} \mbox{for }  (\beta^-)}
                           \end{array}
                           } %  Ende der \arraycolsep - Umgebung
                             \right.
              \quad . \nonumber
   \end{eqnarray}
 %                  ------------------------
\PostScript{6}{0.3}{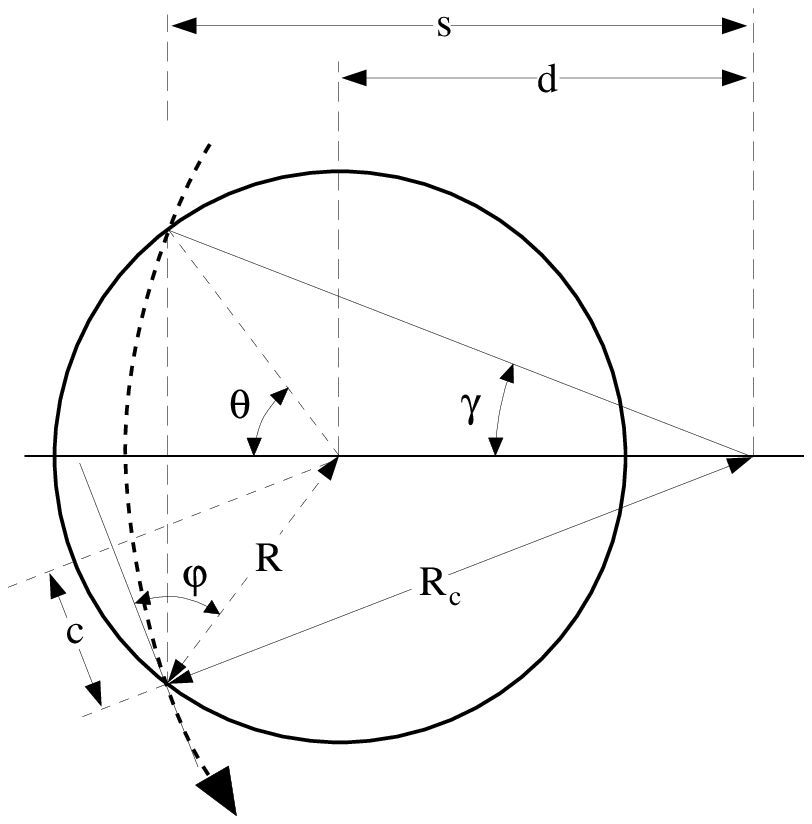}{-0.8}
           {The actions and amplitudes of the classical periodical
            orbits can be expressed purely in terms of the geometrical
            quantities shown here.}
           {geoGR}
%                  ------------------------
  According to the trace formula~\cite{steven:cont}, the orbit amplitudes are
  composed of an integral over the symmetry group, which for the rotational
  $U(1)$ symmetry of the disk just gives $2 \pi / v$, of the period of the
  orbit $L/\hbar k$, and of the Jacobian resulting from the symmetry reduction
  $\mbox{d} L / \mbox{d} \Psi$, where $\Psi=-2n \Theta$. All these quantities can be
  calculated analytically, resulting in
  \begin{eqnarray}
   \label{amps}
   A_\beta&=&\frac{1}{E_0}\frac{1}{\sqrt{Rk\pi}} \frac{1}{\sqrt{v}}
             \frac{R_c}{R} \sqrt{\frac{cd}{sR}} \; \xi_\beta \quad ,
             \\
   \quad \xi_\beta &=&
            \left\{\begin{array}{ll}
                      \pi-\gamma   & \mbox{for } (\beta^+,R_c<R) \\
                      \gamma       & \mbox{otherwise}
                     \end{array}
            \right. \; , \nonumber
  \end{eqnarray}
  where $c,d,$ and $s$ are the geometrical lengths sketched in
  Fig.~\ref{geoGR}. The connection of these geometrical quantities to the
  classification parameter $\beta^\pm$ and the cyclotron radius $R_c$ is given
  in Appendix~\ref{cds}.

%---------------------- Cyclotron orbits -----------------------------
 \subsection{Cyclotron orbits}
  \label{cyc}
  For magnetic fields stronger than $\widetilde B = kR$, the classical
  cyclotron radius $R_c$ is smaller than the disk radius $R$. This gives rise
  to a new class of periodic orbits, the {\em cyclotron orbits}, which do not
  touch the boundary at all (see Fig.~\ref{OrbitsB}). They form translationally
  degenerate families, whereas the bouncing orbits $(v,w)^\pm$ considered above
  are degenerate with respect to rotations. For the translational case, the
  symmetry reduction can be performed directly, without need of the general
  procedure of Creagh and Littlejohn. We transform the phase-space coordinates
  according to
  {\setlength{\arraycolsep}{0.2em}
   \begin{eqnarray}
    \pi_x :=  \frac{1}{\sqrt{|eB|}} \left( p_x+\frac{eB}{2}y \right);
    & &
    \Pi_x  :=  \pi_y+ \sqrt{|eB|}\;x
       \nonumber \\
    \pi_y :=  \frac{1}{\sqrt{|eB|}} \left( p_y-\frac{eB}{2}x \right);
    & &
    \Pi_y  :=  \pi_x- \sqrt{|eB|}\;y  \; . \nonumber \\
   &&
   \end{eqnarray}
  }
  Apart from a factor $\sqrt{|eB|}$, $(\pi_x,\pi_y)$ are the coordinates of the
  motion relative to the center of gyration $(\Pi_x,\Pi_y)$, as illustrated in
  Fig.~\ref{koords}.
%                  ------------------------
\PostScript{4}{0.5}{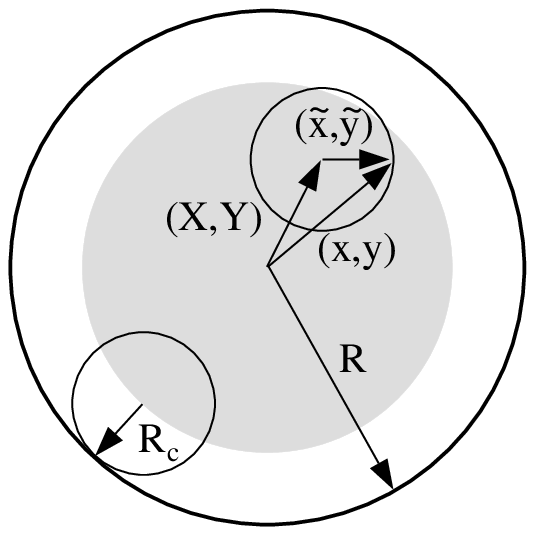}{-0.8}
           {The motion of a charged particle in a homogeneous magnetic
            field can be expressed in the coordinates of the relative
            motion
            $(\tilde x, \tilde y)=|eB|^{-1/2}(-\pi_y,\pi_x)$
            and the coordinates of the center of gyration
            $(X,Y)=|eB|^{-1/2}(\Pi_x,-\Pi_y)$.
            The Hamiltonian is independent of $(\Pi_x,\Pi_y)$;
            all orbits with the center $(X,Y)$ in the gray shaded
            area are degenerate.}
           {koords}
%                  ------------------------
  In these coordinates the Hamiltonian reads
  \begin{equation}
   \label{HO}
   H=\frac{eB}{2m} (\pi_x^2+\pi_y^2) \quad .
  \end{equation}
  As expected, $H$ does not depend on the coordinates of the center of
  gyration. $\Pi_x$ and $\Pi_y$ are canonically conjugate variables, since
  $[\Pi_x,\Pi_y]=i \hbar$. Because the relative and the center-of-gyration
  coordinates commute, i.e.,
  $[\Pi_x,\pi_x]=[\Pi_x,\pi_y]=[\Pi_y,\pi_x]=[\Pi_y,\pi_y]=0$,
  the degeneracy of a cyclotron orbit is simply the phase-space volume $V$
  accessible for $(\Pi_x,\Pi_y)$, which can be directly read off
  Fig.~\ref{koords} (shaded area). We therefore get for the degeneracy
  \begin{equation}
   \label{cyc_degeneracy}
   N=\frac{V}{2 \pi \hbar}
    =\frac{\tilde{B}}{2} \left( 1- \frac{R_c}{R} \right)^2 \qquad .
  \end{equation}
  Now, the Hamiltonian Eq.~(\ref{HO}) is identical to that of a one-dimensional
  harmonic oscillator. Using its analytically known trace formula,\footnote{The
           harmonic oscillator is one of the few cases that can be treated
           {\em exactly} within standard POT~\cite{matthias:oscillator}.}
  the contribution of the cyclotron orbits to the oscillating part of the
  level density is given by
  \begin{equation}
   \label{ZycPOT}
   \delta g^{c}=\frac{1}{2 E_0}\left( 1- \frac{R_c}{R} \right)^2
                          \sum_{n=1}^\infty \cos(nk\pi R_c -n \pi) \quad.
  \end{equation}
  Here $n$ is the winding number around the center of gyration. Note that the
  frequency is again determined by the classical action along the orbit, which
  in this case is
  \begin{equation}
   \label{action_cyc}
   S=n \cdot \hbar k \cdot \pi R_c \qquad .
  \end{equation}
  Note that here exactly half of the kinetic contribution to the action is
  canceled by the flux term.

%------------------------ Additional phases ---------------------------
 \subsection{Additional phases}
  For a discussion of the additional phases in the trace formula
  (\ref{trace_formula}) we refer to Sec.~\ref{Maslov}. There we find that
  the Maslov index for bouncing orbits is $\sigma=3v$, and for the cyclotron
  orbits it is $\sigma=2$. According to~\cite{steven:cont,steven:rest},
  we have an additional phase of $\delta \; \pi/2$ stemming from the
  symmetry reduction. $\delta$ is found to be
  \begin{equation}
   \delta=\left\{ \begin{array}{c@{\hspace{5mm}}l}
                    0 & \mbox{for } (\beta^+, R_c < R) \\
                    1 & \mbox{otherwise.}
                  \end{array}
          \right.
   \qquad .
  \end{equation}
  We have now analytic formulas for all quantities of the trace formula. The
  numerical evaluation of this semiclassical level density will be performed in
  Secs.~\ref{WeakB} and~\ref{StrongB}.

% -------------------- The shell structure --------------------------
\subsection{The shell structure}
\label{shell}
  In an experiment, the observed levels are always broadened due to
  temperature, life-time, or impurity effects. In most systems the levels are
  not approximately equally spaced, but occur in bunches, the so-called {\em
  shells}, which are separated by relatively wide energy gaps. Smoothing the
  level density over a width larger than the typical level spacing, but smaller
  than the distance of the bunches, reveals the {\em (gross-) shell structure}
  of the system. It contains in many cases the dominating quantum effects. This
  folding procedure can easily be implemented in the semiclassical trace
  formula. For {\em pure billiard systems}, i.e., systems where $S=k L$
  with $L$ independent of $k$, a Gaussian folding of the level density
  is equivalent to the multiplication of the orbit amplitudes
  in the trace formula with a Gaussian  with reciprocal width. In
  Appendix~\ref{eval} we give a more general form of this relation, which is not
  restricted to pure billiard systems and to Gaussian smoothing, but can deal
  with general systems and arbitrary smoothing functions. We will in the
  following use this generalized approach, as in finite magnetic fields we no
  longer have a pure billiard system. As this is a more technical point, we
  leave the discussion for Appendix~\ref{eval}. There we also give detailed
  information about the numerical evaluation scheme and the smoothing function
  used. The latter is in the following characterized by a parameter $\tilde
  \gamma$, which  corresponds to the variance of a Gaussian $\exp[-1/2 \left(kR
  / \tilde \gamma  \right)^2] $ with the same half-width.

  The additional factor in the amplitudes stemming from the smoothing strongly
  suppresses the longer periodic orbits,\footnote{This holds only for
               billiard systems. The suitable generalization is again
               given in Appendix~\ref{eval}.}
  so that usually only a few of them (2 - 10) contribute to the gross-shell
  structure. This makes the POT a very convenient tool for the calculation of
  this quantity. The quan\-tum-\-mechani\-cal approach is in some sense complementary
  to the semiclassical one. It first gives the single eigenvalues, of which
  many have to be known to calculate the shell structure. On the other hand, a
  {\em full semiclassical quantization}, i.e., resolving the level density down
  to the single eigenenergies, involves in general an exponentially increasing
  number of orbits and thus is a very demanding task. Here we are mainly
  interested in the semiclassical calculation of the gross-shell structure,
  for which only a few of the shortest and most degenerate periodic
  orbits are required. We will, nevertheless, also try to go for a full
  quantization --- mainly to verify the quality of our semiclassical
  approximation.

%------------------ Results in the weak-field regime ---------------------
\subsection{Results in the weak-field regime}
 \label{WeakB}
 In the previous sections we have derived an analytical trace formula for the
 circular billiard in homogeneous magnetic fields of arbitrary strength.
 Presently we shall discuss the resulting level densities as a function of
 energy and magnetic field. Let us start with weak fields ($R_c > R$),
 for which the topology of the classical periodic orbits is the
 same as in the absence of a magnetic field (see Sect.~\ref{classification}),
 so that we expect the semiclassical approach to be of the
 same quality as for zero field. The high-field regime, where we expect new
 effects to arise, will be the topic of the next section.

 The case of the circular billiard in small homogeneous magnetic fields has
 already been treated by Bogachek and Gogadze~\cite{boga} and by Reimann
 \etal~\cite{Steffi:B} using a perturbative approach for weak fields. Replacing
 the amplitudes of Eq.~(\ref{amps}) by their asymptotic values for
 $\widetilde{B} \to 0$ and expanding the actions of Eq.~(\ref{actions}) up to
 first order in $\widetilde{B}$ reproduces, indeed, their results.

 In Fig.~\ref{CoarseKleinB} the semiclassical level density obtained with
 ${\widetilde \gamma} = 0.35$ (solid line) is plotted against the equivalently
 smoothed quantum result (dashed) for various values of $\widetilde{B}$.
%                  ------------------------
\PostScript{5.4}{0.5}{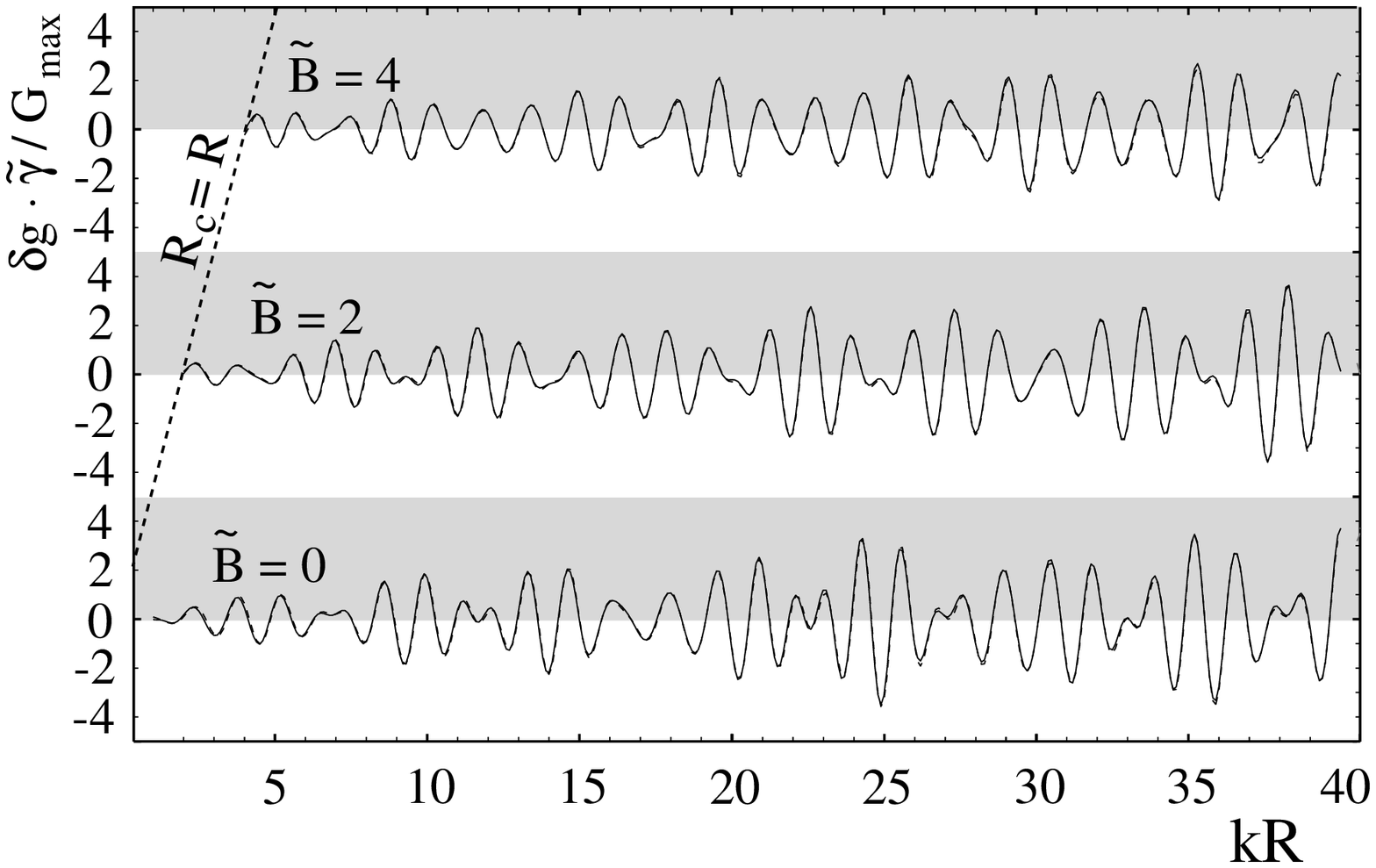}{-0.8}
%\PostScript{9}{0.3}{CoarseKleinB.ps}{-0.5}
           {The semiclassical level density of the disk billiard (solid line)
            matches perfectly the equivalently smoothed quan\-tum-\-mechani\-cal
            result (dashed, and well hidden under the solid line). The smoothing
            width is $\tilde{\gamma}=0.35$.}
           {CoarseKleinB}
%                  ------------------------
 The agreement is almost perfect --- just as it is in the zero-field case, which
 has been extensively discussed by Reimann {\em et
 al.{}}~\cite{Steffi:disk}. Note that
 the calculation requires over 850 numerically determined eigenvalues for the
 quan\-tum-\-mechani\-cal calculation (which then have to be smoothed), whereas the
 semiclassical calculation is analytical and just requires the most
 important orbits (the diameter and the two triangle, square, pentagon and
 hexagon orbits).

 Since we have a classification of all periodic orbits and analytic expressions
 for their actions and amplitudes, we can attempt a full semiclassical
 quantization by summing up sufficiently many of them. The result is shown in
 Fig.~\ref{FullQkleinB}, where we display the {\it total} level density
 $g=\tilde{g}+\delta g$, averaged over a width ${\widetilde \gamma} = 0.025$,
 which is smaller than the typical level spacing. As Tanaka has shown
 in~\cite{kaori}, the smooth part of the level density of the circular billiard
 does not depend on the magnetic field to leading order in $\hbar$. We use the
 Thomas-Fermi level density for zero field, which is identical to the familiar
 Weyl expansion~\cite{BaltesHilf}
 \begin{equation}
  \label{SmoothPart}
  \tilde{g}(k)=\frac{1}{4 E_0} \left( 1- \frac{1}{kR} \right)
  \quad .
 \end{equation}
%                  ------------------------
\PostScript{5.2}{0.5}{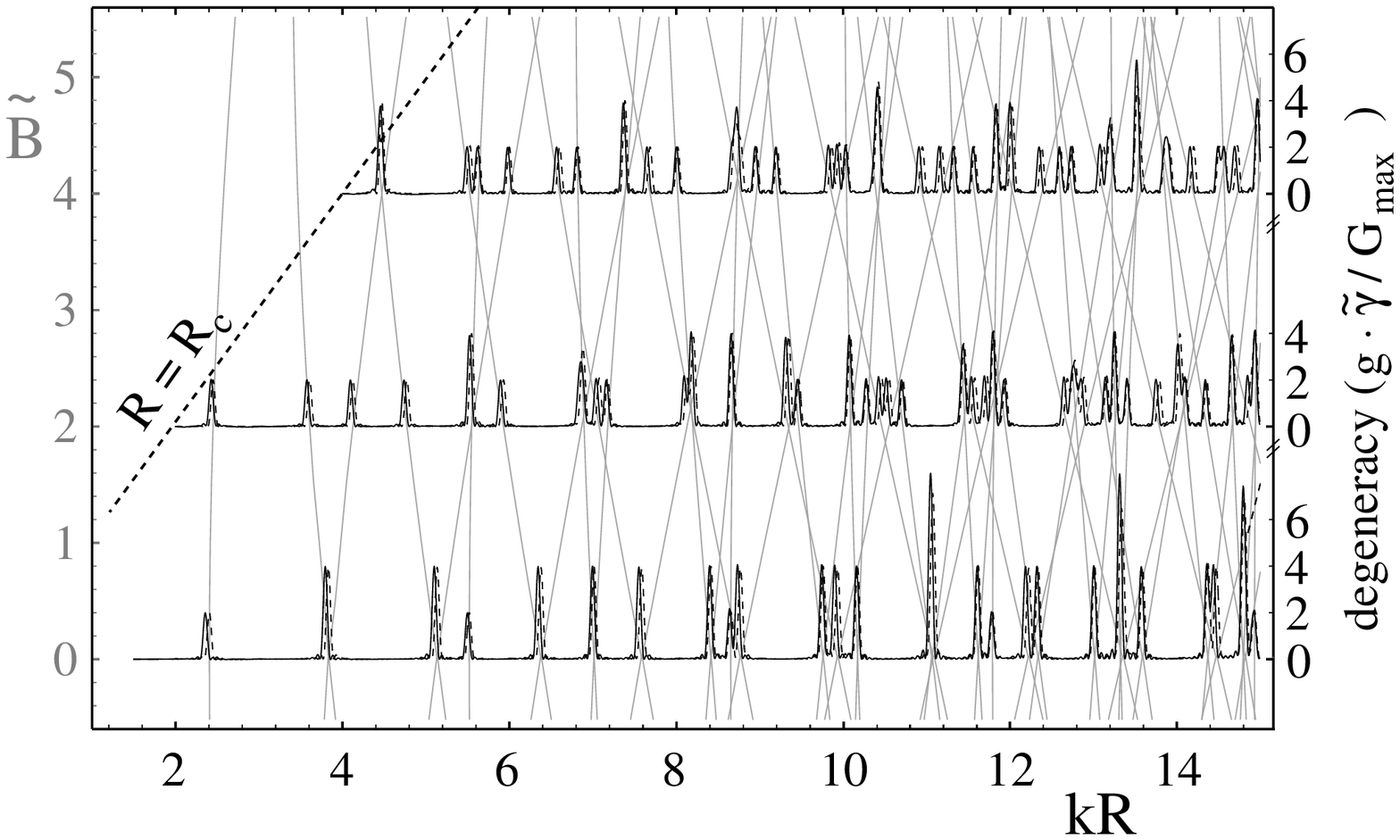}{-0.8}
%\PostScript{9}{0.3}{FullQkleinB.ps}{-0.5}
           {The semiclassical level density (solid) for a smoothing
            width $\tilde \gamma = 0.025$, which is small enough to
            resolve the single eigenenergies (``full quantization'').
            One can hardly distinguish this from the equivalently
            smoothed quan\-tum-\-mechani\-cal result (dashed line underneath).
            The positions of the quan\-tum-\-mechani\-cal eigenvalues in
            dependence of $\tilde{B}$ are indicated by the gray lines.}
           {FullQkleinB}
%                  ------------------------
 Both the semiclassical level density (solid) and the corresponding 
 quantum-mechanical one (dashed) in Fig.~\ref{FullQkleinB} 
 exhibit clearly separated peaks whose heights
% (here rescaled by a factor $\tilde \gamma / G_{\rm max}$)
 give the degeneracies of the individual levels. The two lines can hardly be
 distinguished, thus the semiclassical approach gives almost perfect results
 even in this extreme case of full quantization.

% -------------- Results in the strong-field regime ------------------
 \subsection{Results in the strong-field regime}
  \label{StrongB}

   Figure~\ref{CoarseNoP} is the strong-field equivalent of
   Fig.~\ref{CoarseKleinB}. It displays again the semiclassical
   (solid) and the quan\-tum-\-mechani\-cal (dashed) level densities, obtained with an
   equivalent averaging width $\widetilde \gamma$ = 0.35.
%                  ------------------------
\PostScript{5.2}{0}{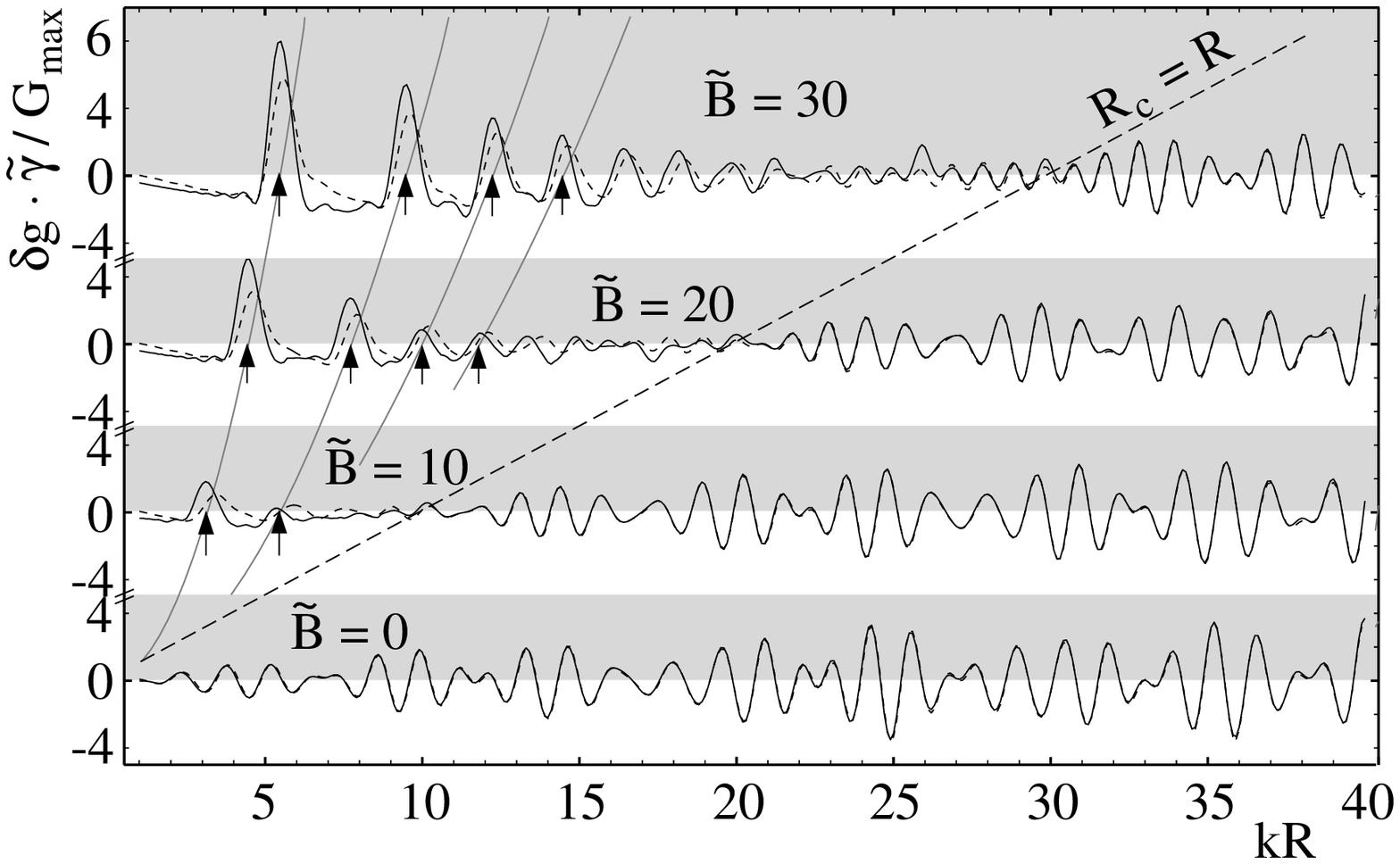}{-0.8}
%\PostScript{9}{0.3}{CoarseNoP.ps}{-0.5}
          {The semiclassical level density ($\tilde{\gamma} = 0.35$) of
           the disk billiard (solid) compared to the equivalently smoothed
           quan\-tum-\-mechani\-cal result (dashed). The gray lines and the arrows
           indicate the positions of the first four Landau levels. For strong
           fields ($R_c < R$), the agreement between the semiclassical and the
           quan\-tum-\-mechani\-cal results is not satisfactory.}
          {CoarseNoP}
%                  ------------------------
   The agreement for small fields ($R_c < R$) is good, as already shown in
   Sec.~\ref{WeakB}. For stronger fields, the positions of the Landau levels
   (gray lines in Fig.~\ref{CoarseNoP}) are well reproduced, but their
   degeneracies are overestimated in the semiclassical approximation.

   Figure~\ref{FullQnoP} corresponds to
   Fig.~\ref{FullQkleinB} and displays the full quantization
   of the system.
%                  ------------------------
\PostScript{5.2}{0.3}{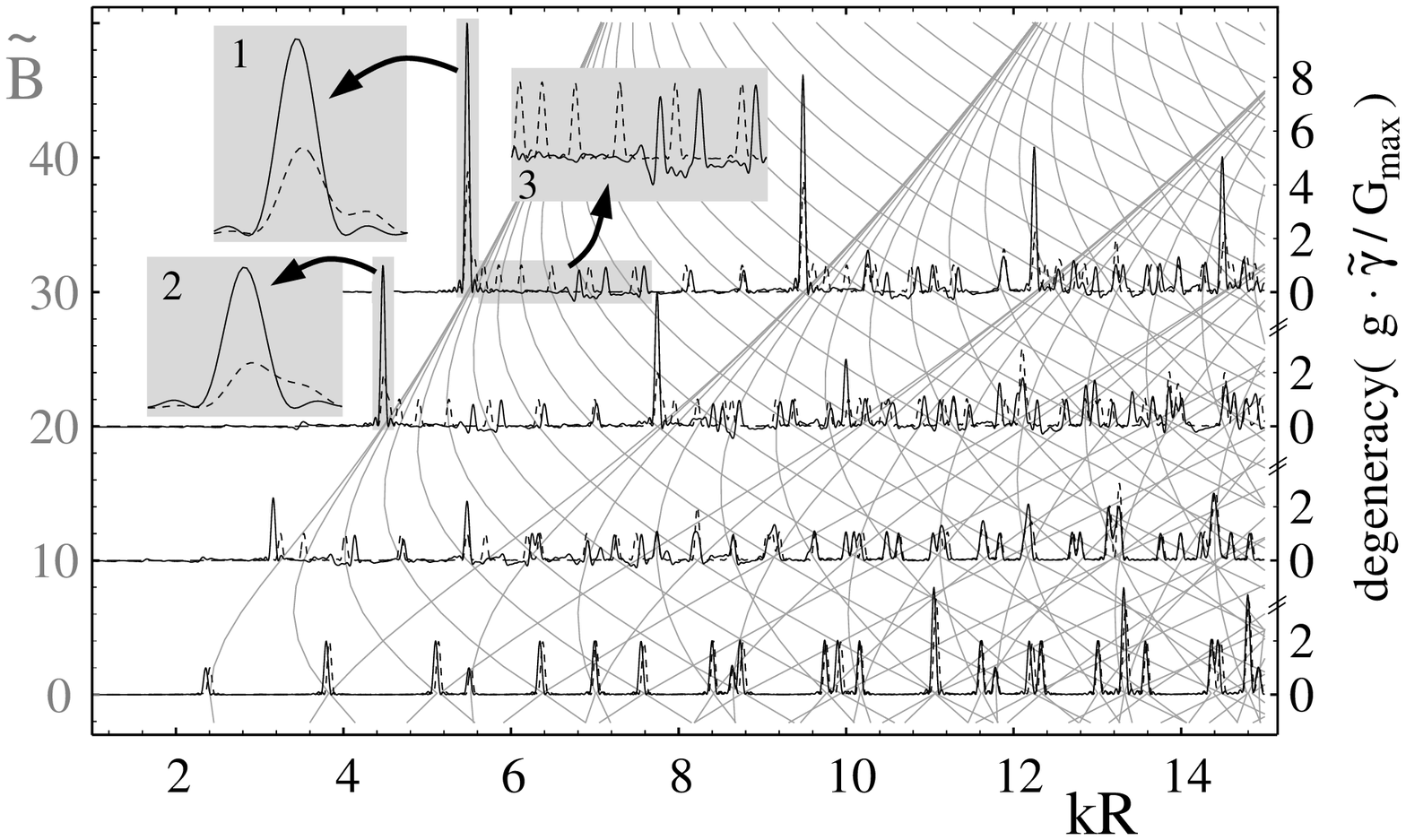}{-0.8}
%\PostScript{9}{0.3}{FullQnoP.ps}{-0.5}
           {The semiclassical level density resolved up to the single
           eigenenergies (solid) compared to the equivalently smoothed
           quan\-tum-\-mechani\-cal levels (dashed). The gray lines indicate the
           positions of the eigenenergies as functions of $\tilde{B}$. For
           strong fields ($R_c < R$) the agreement between semiclassical and
           quan\-tum-\-mechani\-cal calculation is bad. Note especially the large
           discrepancies in the degeneracies of the Landau levels, and the
           completely missing states slightly above the Landau levels
           (see insets).}
           {FullQnoP}
%                  ------------------------
   The semiclassical approach is seen to fail for stronger fields. As already
   mentioned, this is due to the neglect of various effects. First, there are
   orbit bifurcations where classical orbits vanish pairwise with increasing
   magnetic field (see Figs.~\ref{bifurkation}, \ref{Laengen}
   and~\ref{Amplituden}). The change of the topology of the $\beta^+$ orbits
   and the occurrence of cyclotron orbits are also bifurcation effects. Those
   are known to lead to divergences in the trace formula. Second, we have
   neglected boundary effects from grazing orbits and diffraction effects, which
   could be implemented in the trace formula by considering creeping orbits. A
   closer look at Fig.~\ref{FullQnoP} gives some hints as to which of these
   effects dominate. The two most striking observations are as follows:
       (1) The trace formula reproduces well the positions of the Landau
           levels,\footnote{This is no surprise since the Landau levels are due
           to the free cyclotron motion of the electrons, which is equivalent
           to that of a 1D harmonic oscillator. The latter is known to be exact
           in the semiclassical approximation.} but it overestimates the
           degeneracies of these states. The error becomes smaller with
           increasing field strength (see insets in Fig.~\ref{FullQnoP}).
       (2) The levels that have energies slightly above the Landau
           levels are completely missed by the semiclassical approach.
   These two observations suggest that it is a boundary effect that causes the
   discrepancies. A simple hand-waving argument might be useful to illustrate
   the effect. Quantum mechanically, a particle moving on a cyclotron orbit
   will feel the boundary even if classically not touching it. Particles on
   cyclotron orbits close to the boundary thus feel an additional confinement.
   This restriction to a smaller volume will lead to a higher energy. In this
   picture, not all the cyclotron orbits are degenerate. The orbits close to
   the boundary will no longer have the energy of the Landau level, but a
   slightly higher one. This is exactly what would correct the observed defects
   of the semiclassical approximation. In the next section we will present a
   simple way to incorporate this boundary effect in the trace formula.

% ----------- Boundary corrections to the trace formula ---------------
 \section{Boundary corrections to the \\ trace formula}
  \label{Boundary}
  The observations of Sec.~\ref{StrongB} suggest that boundary effects are
  responsible for the failure of the semiclassical approximation in strong
  fields. The only place where boundary properties enter the standard
  trace formula is
  the Maslov index. We therefore propose here to replace the Maslov index by a
  more sophisticated quantity, which includes some quantum effects.
  Before doing this, let us give a brief summary of the origin of the
  Maslov index.

%--------------------------- The Maslov index ----------------------------
  \subsection{The Maslov index}
   \label{Maslov}
   The origin of the Maslov index can most easily be understood in the
   one-dimensional case. Semiclassically, one approximates the wave
   functions to 
   lowest order by plane waves with the local wave number $k(x)=\sqrt{2m
   [E-V(x)]}$. This approximation obviously breaks down at the classical
   turning points where $E=V(x)$ and the wavelength diverges. Expanding the
   wave function around the classical turning points and matching the solutions
   to the plane-wave solutions far from the turning points leads to additional
   phases in the semiclassical quantization \cite{WKB}. In the limit
   $\hbar \to 0$ these are independent of the detailed shape of the potential.
   Each reflection at a soft\footnote{``Soft'' here means that the slopes of the
   potential at the classical turning points are finite.} turning point gives a
   phase of $-\pi/2$, whereas each reflection at an infinitely steep wall gives
   a phase of $-\pi$. Writing this phase as $-\sigma \; \pi/2$, one usually
   calls $\sigma$ the {\em Maslov index}.

   In the case of the circular disk, the Maslov index can be obtained simply by
   counting the classical turning points of the one-dimensional effective
   potential in the radial variable $r$. For skipping orbits, the Maslov index
   per bounce is 3, including one soft reflection at the centrifugal barrier
   and one hard-wall reflection. For the cyclotron orbits, the effective
   potential is a one-dimensional harmonic oscillator (see Sec.~\ref{cyc})
   with two soft turning points, and thus their Maslov index per period is 2.

%----------------------- Reflection phases -----------------------------
  \subsection{Reflection phases}
   \label{Rphase}
   For finite $\hbar$ the additional phase stemming from classical turning
   points will depend on the shape of the potential.
   Let us consider a cyclotron orbit at a distance $x_W$ to the billiard
   boundary. Neglecting the curvature of the boundary (which corresponds to the
   strong-field limit), we can reduce the motion in the 
   presence of the wall to an
   effective 1D motion just as in the unbounded case presented in
   Sec.~\ref{cyc}. This is shown in Fig.~\ref{Eindim}.
%                  ------------------------
\PostScript{3.7}{0}{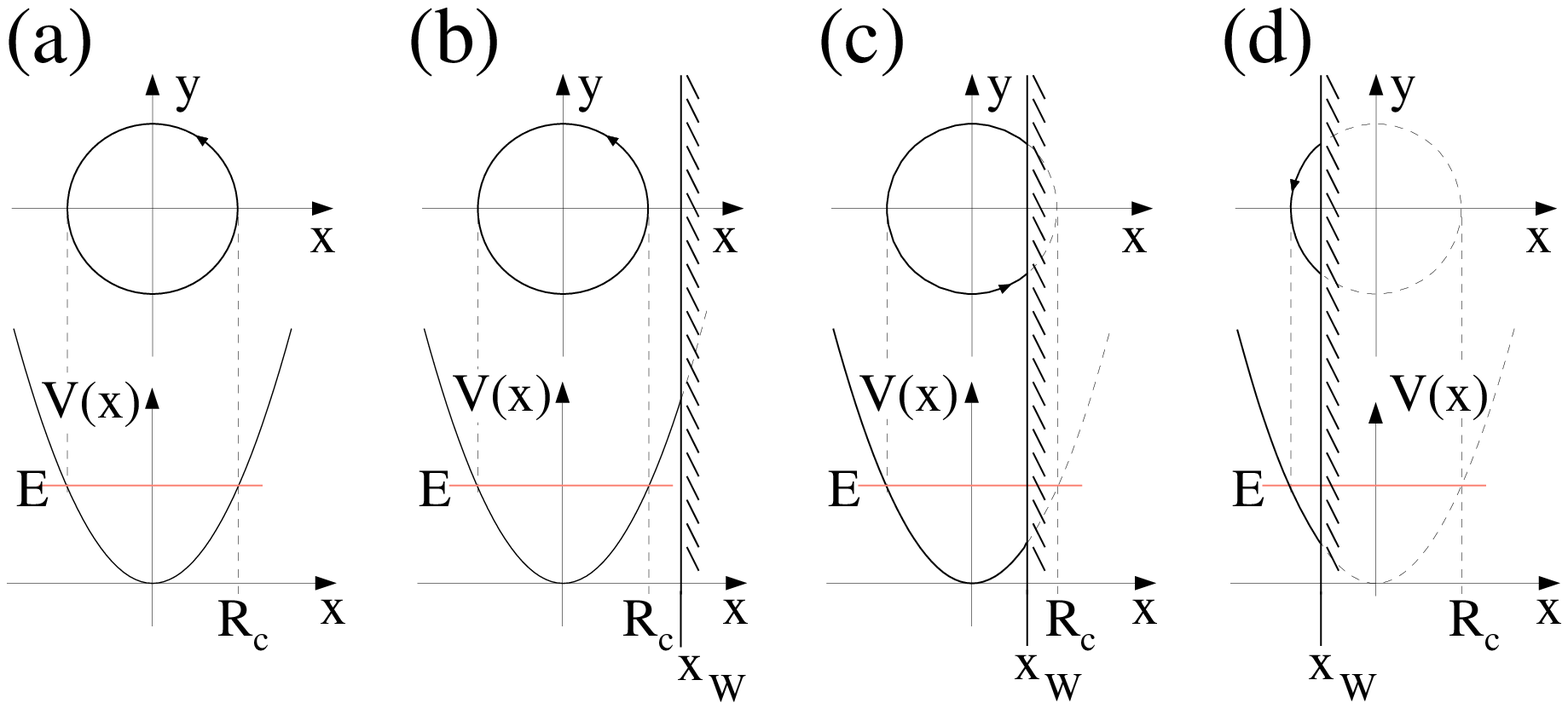}{-0.8}
%\PostScript{6}{0.3}{Eindim.ps}{-0.5}
           {The planar cyclotron orbit is equivalent to the motion in
           a one-dimensional harmonic oscillator (a). Neglecting
           its curvature, the billiard boundary can be
           implemented in the effective one-dimensional motion (b)-(d).}
           {Eindim}
%                  ------------------------
    The upper row of diagrams shows the 2D motion, the lower row gives the
    reduction to the one-dimensional motion in an effective potential.
    Figure~\ref{Eindim} (a) shows the unbounded case, in (b) the orbit is near
    the boundary, and (c,d) illustrate skipping orbits.

    A particle in the potential sketched in Fig.~\ref{Eindim}b is classically
    not influenced by the additional wall, since it will never touch it.
    Quantum mechanically, however, the wave function enters the classically
    forbidden region and thus feels the boundary even for $x_W > R_c$. This
    leads to a smooth transition of the quan\-tum-\-mechani\-cal reflection phase
    $\varphi_R$ over the point $x_W = R_c$, whereas the semiclassical Maslov
    phase is discontinuous at this point; as we have just seen in
    Sec.~\ref{Maslov} above, it is $-\pi$ for $x_W > R_c$ and $-3/2 \pi$ for
    $x_W < R_c$. Our way to implement these quantum effects at the boundary in
    the semiclassical trace formula is therefore to replace the Maslov index by
    the quan\-tum-\-mechani\-cal reflection phase $\varphi_R$ of the corresponding
    one-dimensional motion.
    This smooth version of the Maslov phase will also remove the former clear
    separation between cyclotron orbits and skipping orbits. These two limiting
    cases are now continuously linked, with $\varphi_R$ ranging between $-\pi$
    and $-3/2\pi$. We will refer to the orbits in the transition region, which
    are close to the boundary within $\hbar$, as to the {\em grazing orbits}.

    The calculation of the reflection phases is in this approximation reduced
    to the problem of the one-dimensional harmonic oscillator in an additional
    square-well potential. This system was approached by Isihara and 
    Ebina~\cite{isihara} who used local expansions in terms
    of Airy functions.
    We use a different approach and integrate the 
    quan\-tum-\-mechani\-cal problem
    numerically. From the solutions we calculate the reflection phases
    $\varphi_R$, which is displayed in Fig.~\ref{Phasen}.
%                  ------------------------
\PostScript{5}{0}{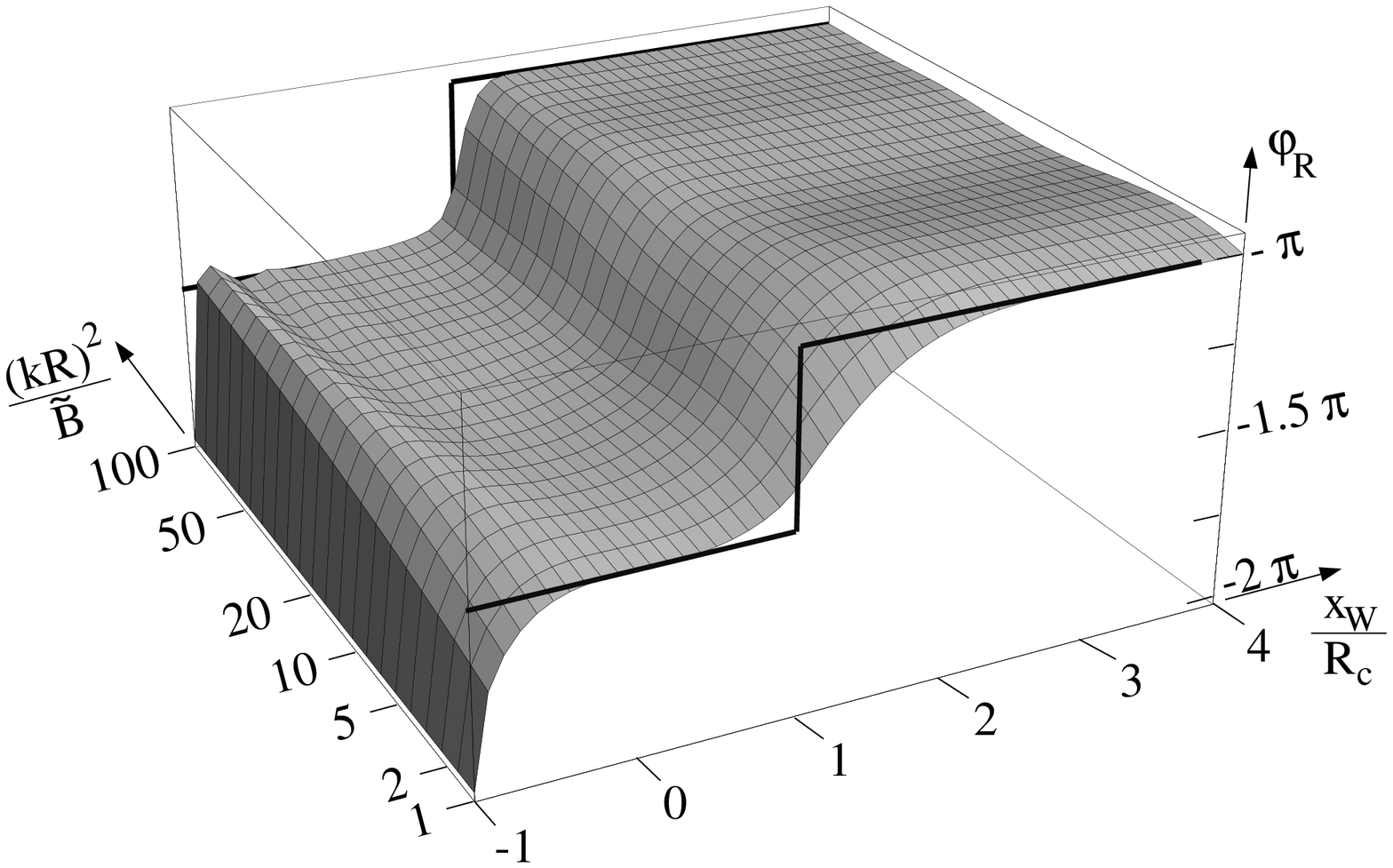}{-0.8}
           {The reflection phase $\varphi_R$ in dependence of the distance of
            the center of gyration from the boundary $x_W$.
            The transition from $x_W <1$ to $x_W > 1$ is continuous and gets
            sharper for increasing $(kR)^2 / \widetilde B$. In the limit
            $(kR)^2 / \widetilde B \to \infty$, which corresponds to the
            semiclassical limit $\hbar \to 0$, the Maslov phase (thick line)
            is recovered.}
           {Phasen}
%                  ------------------------
    As expected, they show a smooth transition from $-\pi$ at $x_W \gg R_c$ to
    $-3/2 \pi$ at $x_W \ll R_c$. The transition gets sharper if
    $(kR)^2/\widetilde B$ increases.
    For $(kR)^2/\tilde B \to \infty$, which corresponds to the
    semiclassical limit $\hbar \to 0$, the standard Maslov phase (thick line)
    is reproduced.
    Quantum corrections are seen to have the greatest influence on the
    grazing orbits $(x_W \approx R_c)$ and on orbits with
    $x_W \mbox{\raisebox{0.5ex}[-0.5ex][0mm]
          {$\scriptstyle >$} \hspace{-1.3em}
          \raisebox{-2ex}[0mm][-2ex]{\Large \symbol{126}}}
    -R_c$.
    These are known as  the {\em whispering gallery}
    orbits, as they move in a narrow region along the boundary.

%-----------Comparison to the quantum-mechanical result-------------------
  \subsection{Comparison to the \\ quantum-mechanical result}
   \label{comp2}
    Figures~\ref{CoarseMitP} and ~\ref{FullQmitP} show the coarse-grained
    level density
    and the full quantization of the spectrum, respectively,
    both calculated with the reflection pha\-ses of Sect.~\ref{Rphase}.
%                  ------------------------
\PostScript{5}{0.5}{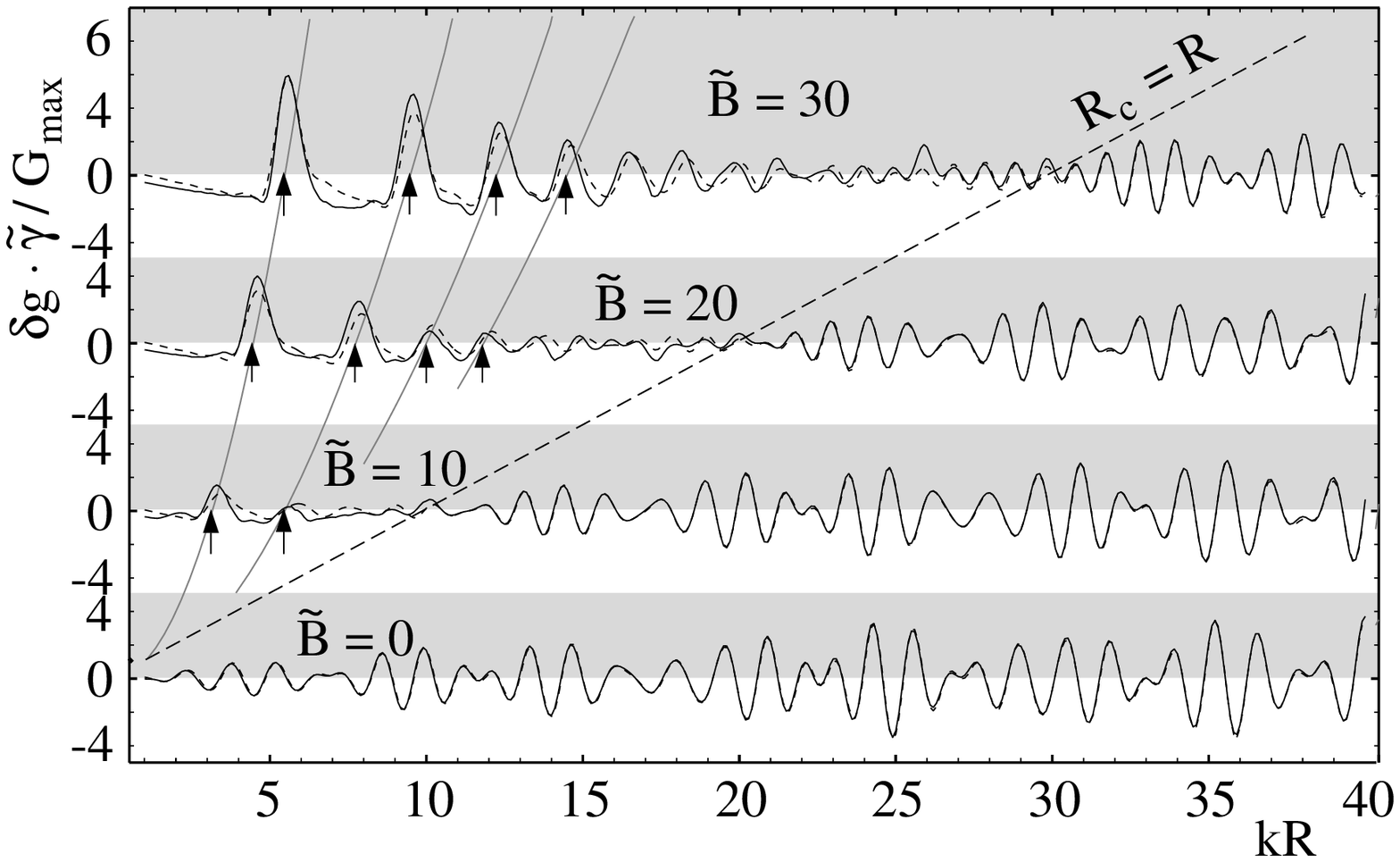}{-0.8}
%\PostScript{9}{0}{CoarseMitP.ps}{-0.5}
           {The semiclassical coarse-grained level density of the disk
           calculated with reflection pha\-ses (solid)
           compared to the equivalently smoothed
           quan\-tum-\-mechani\-cal result (dashed).
           The gray lines and the arrows indicate the positions
           of the lowest Landau levels.
           The agreement is considerably better than with the use of
           the Maslov indices as displayed in Fig.~\ref{CoarseNoP}.}
           {CoarseMitP}
%                  ------------------------
%                  ------------------------
\PostScript{5}{0}{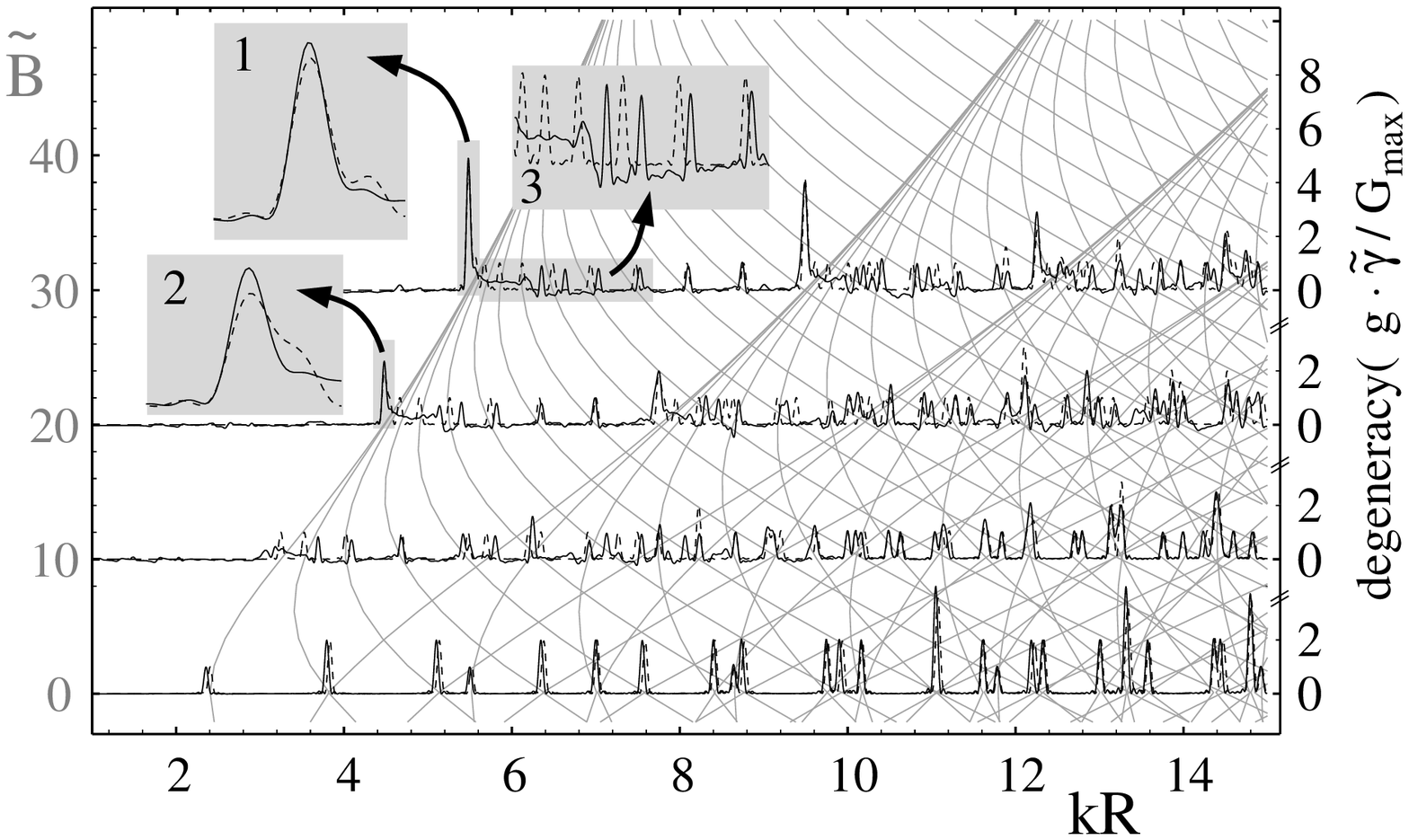}{-0.8}
%\PostScript{9}{0}{FullQmitP.ps}{-0.5}
           {The semiclassical level density with corrected reflection phases
           resolved up to the single eigenenergies of the
           billiard (solid) compared to the equivalently smoothed
           quan\-tum-\-mechani\-cal result (dashed).
           The gray lines indicate the positions of the
           eigenenergies in dependence of $\tilde{B}$.
           The agreement is much better than in the case of the
           Maslov indices in Fig.~\ref{FullQnoP}.
           The degeneracies of the Landau
           levels are correctly reproduced,
           only the levels that are close to
           condensing on the Landau levels show deviations (insets).}
           {FullQmitP}
%                  ------------------------
    A comparison with the corresponding
    diagrams in Figs.~\ref{CoarseNoP} and~\ref{FullQnoP}, which display
    the result obtained with the standard Maslov indices,
    immediately shows that
    the situation is drastically improved when using reflection phases.
    The coarse-grained level
    density now is good at all magnetic field strengths.
    The full quantization displayed in Fig.~\ref{FullQmitP} is not perfect,
    but the most striking error
    in  standard POT, giving
    the wrong degeneracies of the Landau levels, is now corrected.
    This is displayed in detail in the insets 1 and 2 in
    Figs.~\ref{FullQnoP} and~\ref{FullQmitP}, respectively.
    The single states between the Landau levels, however,
    are still not reproduced correctly.
    This is due to our simple approximation, which only includes
    boundary effects via the reflection phase.
    The classical orbits are not changed, so that in our approximation
    the center of gyration of the cyclotron orbits
    ($x_W > R_c$) is
    fixed, whereas for bouncing orbits ($x_W < R_c$)
    it moves around the disk.
    Modeling the expected smooth transition from $x_W > R_c$ to
    $x_W < R_c$ semiclassically would require
    including diffractive orbits that we have neglected here.
    The resulting error can be understood as follows:
    a generic two-dimensional system has two quantum
    numbers, thus requiring two semiclassical quantization condition.
    The free 2D electron gas in a homogeneous magnetic field
    has an additional dynamical symmetry and only one quantum number
    (labeling the Landau level).
    This additional symmetry is broken by the presence of a curved
    boundary.\footnote{A straight boundary does not
             break the symmetry. This is the reason why in this case it
             is possible to reduce the system to one dimension, which
             we have
             exploited in Sec.~\ref{Phasen} for the calculation of the
             reflection phase.}
    This implies that the semiclassical description in terms of cyclotron
    orbits near
    the boundary misses one quantization condition, which is hidden in the
    broken dynamical symmetry. Therefore these orbits give rise to a
    continuous semiclassical (sub) spectrum.
    Bouncing orbits, however, have the correct symmetry and lead to a
    discrete subspectrum.
    This transition can be seen in Fig.~\ref{FullQmitP}.
    On the low-energy side of inset 3, the semiclassical level density shows a
    continuous spectrum stemming from the grazing orbits,
    whereas the quan\-tum-\-mechani\-cal result gives
    quantized levels.
    This error affects mainly the
    fully quantized spectrum; the influence on the gross-shell structure is
    negligible.

    Figure~\ref{CoarseLang} shows once again the
    semiclassical level density calculated with
    reflection phases, now in the whole range from
    zero field to full Landau quantization (solid).
    The comparison with the exact
    quantum result (dashed) shows that the semiclassical approximation
    is in fact valid for {\em arbitrarily} strong fields.
    Small deviations occur only at the bifurcation points
    of the dominating orbits. As already mentioned, we did not include
    the effects of the bifurcations in our calculation.
    The resulting errors are much smaller than the effect of the reflection
    phase, and they are seen to be more important for the
    gross-shell structure than for the full
    quantization.\footnote{This implies that even though the
       amplitudes are diverging, the trace formula can still be used.
       Note, however, that near the bifurcation points the numerical evaluation
       of the trace formula has to be performed with special care, as
       described in Appendix~\ref{eval}.}

%-----------Semiclassical interpretation of the shell structure----------------
 \subsection{Semiclassical interpretation of the \\ shell structure}
  \label{interprete}
  In Sec.~\ref{comp2} we have shown that the semiclassical
  approximation for the level density is valid for arbitrarily strong
  fields. It reproduces the exact quan\-tum-\-mechani\-cal result with a
  remarkably reduced numerical effort. For the
  quan\-tum-\-mechani\-cal calculation shown in
  Fig.~\ref{CoarseLang} about 2500 eigenvalues had to
  be calculated and numerically smoothed for each value of $\widetilde B$,
  whereas the semiclassical result
  is obtained summing the contributions of just 20
  orbits.\footnote{For $R_c > R$ even 10 orbits are sufficient.}
  The most attractive feature of the semiclassical approximation,
  however, is the simple, intuitive picture it gives. Let us now
  exploit this to explain the behavior of
  the shell structure of the disk billiard in terms of classical
  quantities.
%                  ------------------------
\PostScript{11}{0.5}{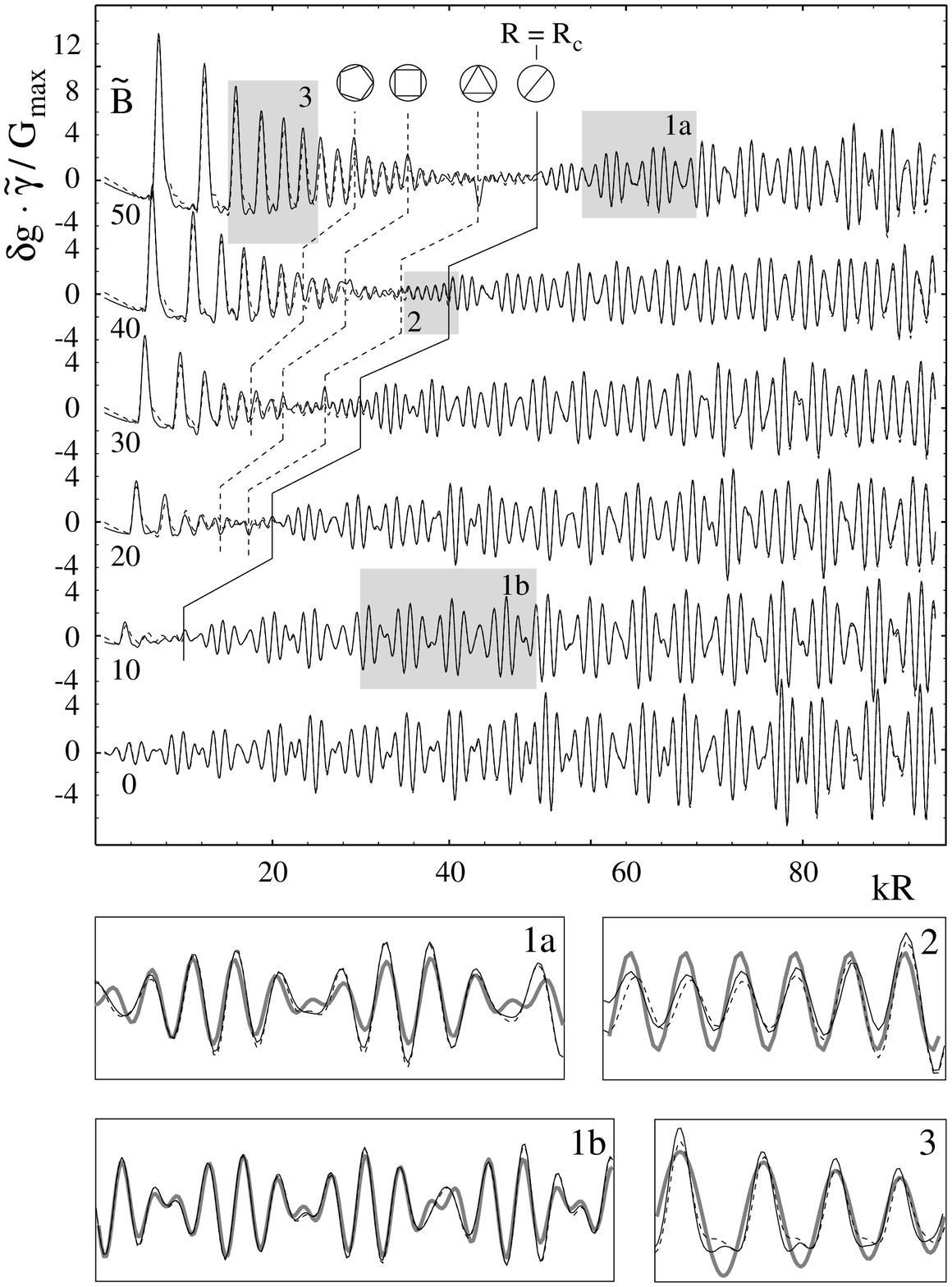}{-0.8}
%\PostScript{15}{0.2}{CoarseLang2.ps}{-0.5}
           {The semiclassical coarse-grained level density of the disk
           billiard with corrected reflection phases (black)
           compared to the equivalently smoothed
           quan\-tum-\-mechani\-cal result (gray).
           The agreement is acceptable in the whole range of energies, disk
           radii, and magnetic fields.
           The vertical lines indicate the bifurcation points of the
           most important orbits.
           The shaded regions are enlarged in the
           figures below. There the thick lines show the
           interpretation of the level density as given in the text.}
           {CoarseLang}
%                  ------------------------
           According to the trace formula Eq.~(\ref{trace_formula}),
           each periodic orbit $\beta$ contributes an oscillating term
           to $\delta g$. Its frequency is determined by the classical
           action $S_\beta$ along this path, which can be locally
           approximated by
           \begin{equation}
            S_\beta(k)=S_\beta(k_0)+\hbar G_\beta(k) \; (k-k_0) \quad,
           \end{equation}
           with the {\em quasiperiod} $\hbar G$. 
           As shown in Appendix~\ref{eval}, 
           for systems with constant absolute velocity along an orbit 
           $G$ is the geometrical orbit length.
           The amplitudes of the oscillating terms are $A_\beta 
           F(G_\beta)$, where $F$ is the window function that depends
           on the desired smoothing of the level density (see
           Appendix~\ref{eval}). Before we interpret the contributions of
           the various orbits to $\delta g$, let us discuss the
           behavior of $G_\beta$ and $A_\beta$.\\ Figure~\ref{Laengen}
           shows the dependence\footnote{The explicit formula for $G$
             is given in Appendix~\ref{eval}.} of $G$ on the ratio
           $R_c/R$.
%                  ------------------------
\PostScript{3.1}{0}{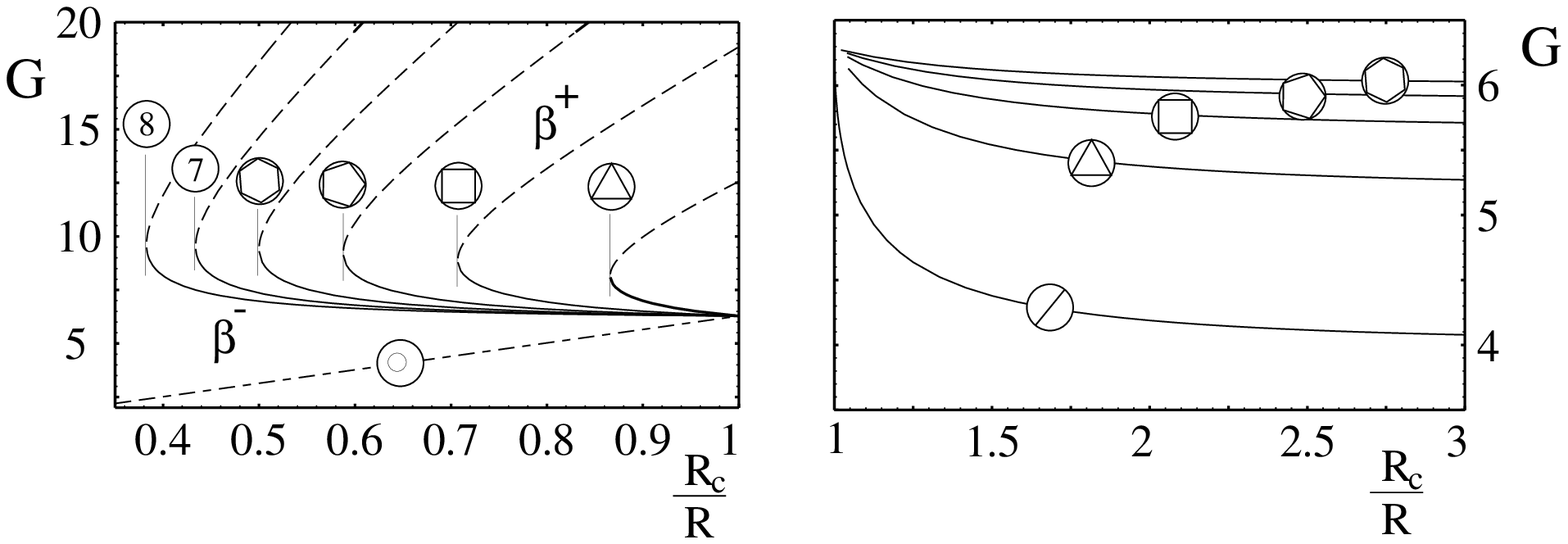}{-0.8}
%\PostScript{6.0}{0.3}{Laengen.ps}{-0.5}
           {The quasiperiods $G$ of the  most important orbits in
            dependence of $R_c/R$. For $R_c > R$, $G$
            is independent of the index $\pm$. 
            The orbit bifurcation points in strong fields
            (vertical lines) can be clearly seen.}
           {Laengen}
%                  ------------------------
  Note that for $R_c>R$ (see right diagram of Fig.~\ref{Laengen})
  $G$ is independent of the direction of motion $\pm$, even if the classical
  action depends on it.
  At $R=R_c$ all orbits are creeping along the boundary, forming 
  collectively the
  {\em whispering-gallery mode}. 
  In strong fields ($R_c < R$, left diagram) $G$ is different for the ``+'' and
  the ``--'' orbits. Only at the bifurcation points, where the two orbits
  coincide, they have identical $G$. For strong fields, the value of
  $G$ at the bifurcation points converges to $w \cdot \pi^2 R$.

  In Fig.~\ref{Amplituden} the amplitudes of the orbits
  relative to the $B=0$ values,
  \begin{equation}
   \label{Anull}
   A_\beta^0 = \frac{\sin ^{3/2} \Theta}{\sqrt{v}},
  \end{equation}
  are plotted versus the ratio
  $kR / \widetilde B = R_c / R$.
%                  ------------------------
\PostScript{5.5}{0.3}{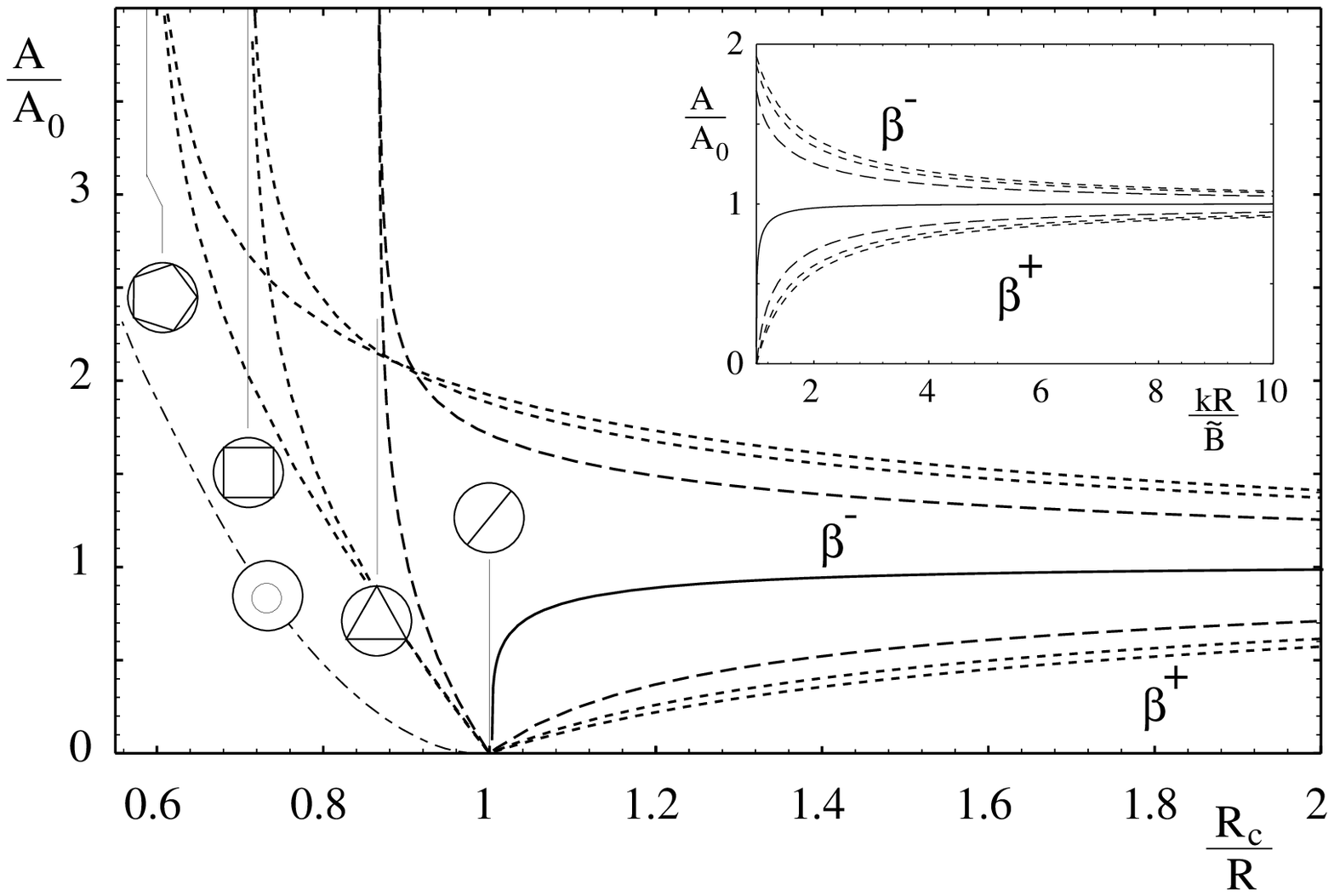}{-0.8}
%\PostScript{8}{0.3}{Amplituden.ps}{-0.5}
           {The amplitudes of the dominating orbits $\beta=(v,1)^\pm$
            with $v=2, \dots, 5$ relative to their $B=0$ value.
            (The amplitude of the cyclotron orbit is in arbitrary units.)
            At the bifurcation points $R_c = \sin (\pi \; v/w)$ indicated
            by vertical lines, the amplitudes diverge.
            For $R_c > R$ the amplitudes of the bouncing orbits quickly
            approach their asymptotic (zero-field) value.
            The inset shows this convergence in a wider range of
            $\widetilde{B}$.}
           {Amplituden}
%                  ------------------------
  The amplitudes of the ``--'' orbit is always larger than that of the
  corresponding ``+'' orbit.
  At $R_c=R$,where the ``+'' orbits change the topology
  (see Fig.~\ref{OrbitsB}), their amplitudes are zero, so that
  these bifurcations do not lead to artifacts in the level density.
  In stronger fields, the amplitudes diverge at the bifurcation points,
  indicating that the semiclassical approximation breaks down
  at these points (more
  exactly, one of the saddle-point approximations in the derivation of
  the trace formula becomes invalid).
  A rigorous treatment of these bifurcations has been
  presented in a very general form for two-dimensional systems
  by Ozorio de Almeida and Hannay~\cite{ozo}, more explicit calculations have
  been performed, for example, by Kus {\em et al.{}}~\cite{kus} and
  Sieber~\cite{sieber}.
  The main idea is always to replace the saddle-point approximation by a
  better adapted uniform integration.  An application to the disk billiard
  has not been attempted here and will be the subject of further studies.

  For the interpretation of the shell structure,
  let us first look at the weak-field regime ($R_c > R$).
  The amplitudes for zero field given in Eq.~(\ref{Anull}) are
  proportional to
  $v^{-1/2}$, favoring orbits with a small number of bounces $v$. The
  dependence of the amplitudes on the magnetic field as shown in
  Fig.~\ref{Amplituden} indicates that in the region where the
  ``--'' orbits differ significantly from the ``+'' orbits, the
  latter are negligible. These
  effects\footnote{The $G$ dependence of $F(G)$
          also supports slightly this effect.}
  together strongly favor the
  $(2,1)$ and the $(3,1)^-$ orbits. They end up with comparable
  amplitudes.
  From this picture we expect as the dominating feature of the
  level density a pronounced beating pattern
  from the interference of the diameter and the triangular orbit.
  This beating pattern is well known for the zero-field case.
  In three-dimensional metal clusters, it is usually referred to as
  {\em supershell oscillations}
  \cite{nishioka}.\footnote{In the 3D spherical cavity, the beat
     comes from the interference of the
     triangle and the square orbits (see Ref.~\cite{BB2}).}
  Our description suggests that this
  beating will survive in homogeneous magnetic fields up to a
  strength of $\widetilde{B}=kR$. This is
  indeed observed, as seen in Fig.~\ref{CoarseLang}. The
  thick lines in the frames~(1a) and~(1b) correspond to
  a function\footnote{The phases are, of course, adjusted.}
  \begin{eqnarray}
    \sin(k G_{(1,2)})+\sin(k G_{(1,3)^-})= \nonumber \\
    \hspace{2cm} \sin\left(k  \frac{\Delta G}{2} \right) \;
                  \sin\left(k \frac{\bar G}{2} \right) \quad .
  \end{eqnarray}
  It predicts correctly the structure of the level density in this
  regime.\\

  Approaching the field strength where $R_c=R$, all orbits change $G$ sharply
  to $2\pi$, so that they interfere coherently, forming the whispering
  gallery mode. We therefore expect that the beating behavior will disappear,
  leaving just a simple oscillation with the common frequency.
  In Fig.~\ref{CoarseLang} this sudden stop of the
  beat at $R_c=R$ can be clearly seen. The solid line in frame~2 shows that
  the frequency of the remaining single oscillation is predicted correctly.\\

  For $R_c<R$, the influence of the
  cyclotron orbits increases with stronger fields.
  The large amplitudes of the bouncing orbits near the bifurcation
  points is, as we have
  already pointed out, unphysical and should be removed by a rigorous
  treatment of the orbit bifurcations.
  For strong fields, only cyclotron orbits and bouncing orbits with a
  great number of bounces $v$ exist. The amplitudes of the latter are
  proportional to $v^{-1/2}$, so that in very
  strong fields we expect that the cyclotron orbits dominate the
  level density.
  The gray lines in frame~3 of Fig.~\ref{CoarseLang}
  show the corresponding oscillating
  term,\footnote{For a simpler comparison, the amplitude is chosen to
         rise quadratically, as indicated by
         Eq.~(\ref{cyc_degeneracy}).}
  which, indeed, reproduces the main feature of the
  quan\-tum-\-mechani\-cal result (solid black).
  The skipping orbits with greatest amplitudes are
  those close to their bifurcation points. As can be seen in
  Fig.~\ref{Laengen}, all those orbits have nearly the same value of
  $G=w \cdot \pi^2 R$. Their
  contributions should therefore interfere constructively, giving rise
  to small structures in the level density of this period. Such
  structures can indeed be observed in a higher-resolution spectrum, as
  shown in Fig.~\ref{FullQmitP}. The spacing of the small
  peaks between the Landau levels is, indeed, consistent with our simple
  picture.\footnote{This holds for the spacing of levels that ``belong''
              to the same Landau level, and as long as we
              still have skipping orbits and do not enter the grazing
              orbit regime, where the reflection phases change.}

  Altogether we could show that
  this simple semiclassical picture is able to explain the main
  features of the quite complicated behavior of the level density
  for arbitrarily strong fields in terms of just 3 classical
  periodic orbits.
  We have here interpreted the dependence of the level density
  on the energy, but
  a completely analogous approach for the dependence on the magnetic
  field is possible.

%-------------------Summary---------------------
 \section{Summary}
 \label{sum}
 We have derived a trace formula for the oscillating part of the level
 density of a circular billiard in homogeneous magnetic fields.
 We have used the general approach of Creagh and Littlejohn and compared
 our findings  with the quan\-tum-\-mechani\-cal solution.
 In the weak-field domain, where the classical cyclotron radius
 $R_c$ is larger than the disk radius $R$, the agreement is
 excellent and even a
 full quantization, e.g., the resolution of the level density
 into individual energy levels, is possible.\\
 In stronger fields, the quality of the standard semiclassical approximation
 is not satisfactory, even for the gross-shell structure.
 We have identified boundary effects to be responsible
 for the major part of the deviations.
 To implement these effects in the semiclassical trace formula,
 we have replaced the (discrete) Maslov index by a (continuous) reflection phase.
 The latter was calculated in a simple one-dimensional approximation.
 With this correction,
 the semiclassical approximation to the exact
 quan\-tum-\-mechani\-cal level density is good for all field strengths and energies.
 For a correction of the remaining deviations
 it would be necessary to include diffractive
 orbits and the effects of the orbit bifurcations.
 The orbit bifurcations at strong field strengths affect the shell
 structure only to a small extent, their influence on the full
 quantization is even smaller. The diffractive orbits do not influence the
 shell structure but only the full quantization.
 Both effects can therefore
 be neglected for the semiclassical description of the gross-shell
 structure. The reflection phases, however, are a crucial correction
 for both the gross-shell structure and the full quantization.

 One advantage of the semiclassical description is
 its easy numerical evaluation. Much more attractive, however, is
 the simple, intuitive picture gained from it.
 Quantum mechanics readily gives information on individual
 levels or level statistics, which are hard to
 derive semiclassically. But the experimentally important
 long-range correlations of levels,
 leading to shells and supershells, are very easy to explain
 semiclassically.
 For a qualitative description of the shell structure
 just one or two classical periodic orbits are sufficient.
 In strong fields the single oscillation of the cyclotron orbits
 dominates and the coherent superposition of the strongest skipping
 orbits gives
 rise to additional small structures with much smaller spacing.
 For field strengths  with $R_c \mbox{\raisebox{0.5ex}[-0.5ex][0mm]
                  {$\scriptstyle <$} \hspace{-1.3em}
    \raisebox{-2ex}[0mm][-2ex]{\Large \symbol{126}}} R$ the skipping orbits
 form coherently the whispering gallery mode, which gives rise to a single
 oscillation of the level density. In weak fields, the interference between the
 diameter and the triangular orbit dominates the level density. A quantitative
 semiclassical description is already possible including between 10 and 20
 orbits. Future studies will be aimed at a rigorous treatment of the orbit
 bifurcations and an implementation of diffractive effects.\\
 \vspace{0.5cm}

 {\bf Acknowledgments}\\
 It is a great pleasure to thank Stephen Creagh for his continued interest
 in this work. He inspired many of the ideas presented here, helped to shape
 them in discussions, and gave valuable comments on the manuscript. We also
 thank Stephanie Reimann for stimulating discussions. This work has been
 supported by the Commission of the European Communities under Grant No.
 CHRX-CT94-0612.

%--------------------APPENDIX-------------------------------
\begin{appendix}
\section{Geometrical quantities of the PO}
 \label{cds}
 The geometrical lengths $c$, $d$, and $s$ and the angles $\Theta$, $\gamma$, and
 $\varphi$ sketched in Fig.\ref{geoGR} can be expressed
 in terms of the classical cyclotron radius $R_c$, the disk radius $R$ and the
 classification parameter $\beta^\pm=(v,w)^\pm$ as follows:
 \begin{eqnarray}
  \Theta  & = & \frac{w}{v}\pi \quad, \nonumber \\
  \gamma  & = & \arcsin \left( \frac{R}{R_c} \sin \Theta \right)
                \quad, \nonumber \\
  \varphi & = & \left\{ \begin{array}{rcccll}
                         \gamma &  - &\Theta &+ & \pi/2 &
                               \mbox{for } (\beta^+, R_c > R) \\
                         -\gamma & + &\Theta &+ & \pi/2 &
                               \mbox{for } (\beta^+, R_c < R) \\
                         \gamma & + &\Theta &- & \pi/2 &
                               \mbox{for } (\beta^-) \quad,
                        \end{array}
                \right. \nonumber \\
  c & = & R \cos \varphi \quad, \nonumber \\
  s & = & \sqrt{{R_c}^2-R^2 \sin ^2 \Theta } \quad, \nonumber \\
  d & = & \left\{  \begin{array}{ll}
                    |  s-R \cos \Theta | & \mbox{for }  \beta^+ \\
                    \, s+R \cos \Theta   & \mbox{for }  \beta^- \quad.
                   \end{array}
          \right.
 \end{eqnarray}

% -------------------- Evaluating the PO sum --------------------------
\section{Evaluating the PO sum}
\label{eval}
 Semiclassical trace formulas are asymptotic series with non trivial
 convergence properties, so that they cannot be summed up straightforwardly.
 Frequently the Gaussian smoothing technique is used, which approximates the
 level density folded with a Gaussian by the trace formula where the amplitudes
 are damped by an additional (Gaussian) factor. This approach is limited to
 Gaussian line shapes and to smoothing of the level density in $k$. In this
 Appendix we introduce a more general approach, which can deal with arbitrary
 line shapes and smoothing variables. We will also state explicitly the
 conditions for the approximation to be valid.\\

 The general form of a trace formula is given by
 \begin{equation}
   \label{POT:general}
   \delta g
   \; =\;
   \sum_\Gamma A_\Gamma(E)
        e^{i \frac{S_\Gamma(E)}{\hbar}-i\sigma_\Gamma \frac{\pi}{2}}
   \quad ,
 \end{equation}
   where $\Gamma$ is a one-dimensional classification
   of the classical periodic orbits.
   If there is a generalized energy $e(E)$, and functions $G(\Gamma,E)$
   and $\tilde{\sigma}(G)$, which fulfill
 \begin{equation}
  \label{separation}
  \frac{S_\Gamma(E)}{\hbar}-\sigma_\Gamma\frac{\pi}{2}
  =
  e G-\tilde{\sigma}(G)
  \quad ,
 \end{equation}
 we can rewrite the trace formula as
 \begin{equation}
  \label{FT-summe}
   \delta g \; = \; \sum_{G} A_2(e,G)e^{ieG} \quad .
 \end{equation}
 Rescaling $G$ we can always obtain $G \in {\rm I\kern-.18em N}$; the rescaling factors
 should be included in $A_2(e,G)$.
 Let us first assume that $A_2$
 factorizes in
 terms that only depend on the generalized energy $e$ and the
 classification variable $G$:
 \begin{equation}
  \label{sepAmp}
  A_2(e,G)=A_G(G)\; A_e(e) \quad.
 \end{equation}
 Approximating Eq.~(\ref{FT-summe}) by an integral
 \begin{equation}
  \label{FT-integral}
   \delta g \; \approx \; A_e(e) \int_{G} A_G(G)e^{ieG} \mbox{d} G \quad .
 \end{equation}
 gives (apart from normalization constants) the oscillating part of
 the level density $\delta g$ as the Fourier transform
 of $A_G(G)$:
 \begin{equation}
  \delta g(e) \approx \sqrt{2\pi} \; A_e(e) \; {\cal F}[A_G(G)] \quad .
 \end{equation}
 For an arbitrary {\em window function} $F(G)$ we get,
 using the well-known folding theorem,
 \begin{equation}
  \int_{G} F(G) A_2(e,G) e^{ieG} \mbox{d} G \approx
  \delta g(e) \ast f(e) \quad .
 \end{equation}
 Here $f(e)$ denotes the Fourier transform of $F(G)$ and ``$\ast$''
 stands for the convolution integral.
 Therefore we have
 \begin{equation}
  \label{Faltungssatz}
  \delta g^F :=
  \sum_{\Gamma} F(G) A_\Gamma(E) \;
        e^{i \frac{S_\Gamma(E)}{\hbar}-i\sigma_\Gamma \frac{\pi}{2}}
  \approx
  \delta g (e) \ast f(e) \; ,
 \end{equation}
 where $\delta g^F$ denotes the trace formula with damped amplitudes.
 This relation shows that folding the
 semiclassical level density with a {\em smoothing function} $f(e)$ is
 equivalent to a multiplication of the
 amplitudes with a window function $F(G)$.
 Unfortunately the restrictions
 of Eqs.~(\ref{separation}) and~(\ref{sepAmp}) are quite severe and often
 prevent the application of Eq.~(\ref{Faltungssatz}).
 With two additional approximations we can relax these restrictions.
 In the general case Eq.~(\ref{sepAmp}) is violated and we may just
 separate out  a common dependence of the amplitudes on $e$:
 \begin{equation}
  A_2(e,G)=A_G(e,G) \; A_e(e) \quad .
 \end{equation}
 In this case Eq.~(\ref{Faltungssatz}) is still a good
 approximation if $A_G(e,G)$ is sufficiently slowly varying in $e$.
 If we denote the characteristic width of $f(e)$ with $\gamma$,
 this means that $A_G(e,G)$ has to be nearly constant over a region
 $\gamma$ in $e$.
 If, on the other hand, there are no functions $e(E)$ and $G(E,\Gamma)$ that
 fulfill Eq.~(\ref{separation}), a local expansion of the action $S$ in
 powers of $e$ can be used:
 \begin{equation}
  \label{expansionS}
  \frac{S}{\hbar} =
   \frac{S(e_0)}{\hbar}+G(e_0)\; (e-e_0)+ {\cal O}(e-e_0)^2 \quad.
 \end{equation}
 If this approximation is good in a region in $e$ wider than the
 typical width $\gamma$ of the smoothing function,
 Eq.~(\ref{Faltungssatz}) still holds.
 In the general case $G$ is therefore given by the
 first derivative of the classical action with respect to $e$:
 \begin{equation}
  \label{period}
  G(E)=\left. \frac{1}{\hbar} \frac{\mbox{d} S}{\mbox{d} e} \right|_E  \quad .
 \end{equation}
 With $e=E$, $\hbar G$ is the period $T$ of the orbit, so that
 we refer to $\hbar G$ as the {\em quasiperiod}. 
 For systems with constant absolute velocity along the orbit
 (this holds especially for billiards),
 we get for the choice $e=k$
 \begin{displaymath}
  \frac{\mbox{d} S}{\mbox{d} e}=
  \frac{\mbox{d} S}{\mbox{d} E}\frac{\mbox{d} E}{\mbox{d} k}=
  T \cdot \frac{k \hbar^2}{m}=\hbar L \quad,
 \end{displaymath}
 where $L$ is the geometrical orbit length. 
 Putting all approximations together, we have shown that
 damping the amplitudes in the trace formula with a window function
 depending on $G$ gives an
 approximation for the level density folded with the Fourier transform
 of the window function used:
 \begin{equation}
  \label{resultFT}
  \delta g^F \approx f(e) \ast \delta g \quad .
 \end{equation}
 This is the main result of this Appendix.
 The approximation holds if in a region wider than the typical
 width $\gamma$ of the smoothing function the conditions
 \begin{equation}
  \label{condition1}
%    \frac{S}{\hbar}  \approx
%     \frac{S(e_0)}{\hbar}+G(e_0)\cdot (e-e_0)
    S \approx S(e_0) +G(e_0)\; (e-e_0)
 \end{equation}
 and
 \begin{equation}
  \label{condition2}
     A_2(e,G)    \approx    {\rm const}
 \end{equation}
 are fulfilled. These conditions depend mainly on the behavior of the
 actions and amplitudes. In order to match them, a well-adapted choice
 of the generalized energy is essential. Note that for narrow
 smoothing functions (small $\gamma$),
 the conditions are less restrictive.
 Therefore for a full quantization the use of Eq.~(\ref{resultFT})
 is often justified, whereas for the calculation of the gross-shell
 structure the conditions Eqs.~(\ref{condition1}) and~(\ref{condition2})
 put tight limits on the use of the amplitude damping ansatz
 -- which might seem counter-intuitive at first sight.

 We now illustrate the result with a simple example.
 Pure billiard systems are those where the the action along the orbits
 scales with the wave
 number: $S = \hbar k \cdot L$, and $L$, the geometric orbit length,
 is independent of the energy. Setting
 \begin{equation}
  e(E)=k=\sqrt{\frac{2mE}{\hbar^2}} \quad
  {\rm and} \quad
  G(\Gamma)=L \quad ,
 \end{equation}
 Eq.~(\ref{condition1}) is fulfilled trivially.
 If Eq.~(\ref{condition2}) is also matched, then the use of a window
 function $F$ depending on the orbit length $L$ is equivalent to a folding
 of the level density in $k$. Using a Gaussian window function we get a
 Gaussian smoothing of the level density in $k$ space.
 This is the technique frequently applied
 when evaluating trace formulas for billiard systems.
 Equation~(\ref{resultFT}) is somewhat more general, as it is not
 restricted to billiard systems nor to special window functions.
 It makes (at least in principle) the calculation of arbitrary
 line shapes within the POT possible. It can also be used for
 an estimation of the effects of a (numerical) truncation of
 the trace formula, which can be
 thought of as a special window function.
 More important, however, are Eqs.~(\ref{condition1})
 and~(\ref{condition2}), which give the limits of validity
 of the amplitude damping formula~(\ref{resultFT}).\\

 %--------------- Evaluation for the circular billiard -----------------------
 \subsection{Evaluation for the circular billiard}
  We want to apply the  considerations of the last section
  on the circular billiard.
  The natural choice
  for the generalized energy is $k$. 
  Then the quasiperiod is the geometrical orbit length, given by
  \begin{equation}
   G=vr \cdot \left\{
      {\setlength{\arraycolsep}{0.1em}
       \begin{array}{cccp{3mm}l}
        2\pi & - & 2\gamma 
             & & \mbox{for} \; (\beta^+, R_c < R) \\
             &   & 2\gamma 
             & & \mbox{otherwise.}
       \end{array}
      }
     \right. \quad 
  \end{equation}
  Note that for $R_c>R$ (weak fields) $G$ is independent of the
  direction of motion $\pm$. \\
  For computing the trace formula we have to choose an appropriate
  window function. As we want to compare the semiclassical result with
  the exact quan\-tum-\-mechani\-cal one, we look for a
  window function that can be Fourier transformed analytically.
  The usual Gaussian is nonzero for all $G$ and
  has to be truncated, being thus no longer
  analytically Fourier transformable.
  We used a triangular window instead, which matches all our demands.
  In order to make our results comparable with the usual Gaussian
  smoothing, we characterize the window function with a parameter
  $\tilde{\gamma}$, which  corresponds to the variance of a Gaussian
  $\exp [-1/2 (k/ \tilde{\gamma})^2] $ with the same half-width.\\

  We still have to check if the conditions~(\ref{condition1})
  and~(\ref{condition2}) hold.
  They depend on the behavior of the amplitudes
  that are plotted in Fig.~\ref{Amplituden}.
  At the bifurcation points the orbit amplitudes
  diverge, so that Eq.~(\ref{condition2}) is violated.
  For the evaluation in the corresponding regions we have therefore used
  a numerical folding procedure and evaluated directly the right-hand-side of
  Eq.~(\ref{resultFT}).\\
  For the cyclotron orbits discussed in Sec.~\ref{cyc} we get
  $ G=n \cdot 2\pi R_c $
  and
  $A=(2E_0)^{-1} (1-R_c/R)^2$,
  which is slowly varying in the whole energy range.
  For the cyclotron orbits, approximation~(\ref{resultFT})
  is therefore
  justified for all $\widetilde E$ and $\widetilde B$.
\end{appendix}

% --------------------------------thebibliography----------------------

%**************************************************************************
\end{document}